\documentclass[aip,jcp,reprint,groupedaddress]{revtex4-2}%
\usepackage{graphicx}
\usepackage{tabularx}
\usepackage{dcolumn}
\usepackage{longtable}
\usepackage{tensor}
\usepackage{placeins}
\usepackage{bm}
\usepackage{amsmath}
\usepackage{amsfonts}
\usepackage{amssymb}
\usepackage{xcolor}
\usepackage{float}%
\providecommand{\U}[1]{\protect\rule{.1in}{.1in}}
\begin{document}
\title{Family of Gaussian wavepacket dynamics methods from the perspective of a
nonlinear Schr\"{o}dinger equation }
\author{Ji\v{r}\'i Van\'i\v{c}ek}
\email{jiri.vanicek@epfl.ch}
\affiliation{Laboratory of Theoretical Physical Chemistry, Institut des Sciences et
Ing\'enierie Chimiques, Ecole Polytechnique F\'ed\'erale de Lausanne (EPFL),
CH-1015, Lausanne, Switzerland}
\date{\today}

\begin{abstract}
Many approximate solutions of the time-dependent Schr\"{o}dinger equation can
be formulated as exact solutions of a nonlinear Schr\"{o}dinger equation with
an effective Hamiltonian operator depending on the state of the system. We
show that Heller's thawed Gaussian approximation, Coalson and Karplus's
variational Gaussian approximation, and other Gaussian wavepacket dynamics
methods fit into this framework if the effective potential is a quadratic
polynomial with state-dependent coefficients. We study such a nonlinear
Schr\"{o}dinger equation in full generality: we derive general equations of
motion for the Gaussian's parameters, demonstrate the time reversibility and
norm conservation, and analyze conservation of the energy, effective energy,
and symplectic structure. We also describe efficient, high-order geometric
integrators for the numerical solution of this nonlinear Schr\"{o}dinger
equation. The general theory is illustrated by examples of this family of
Gaussian wavepacket dynamics, including the variational and nonvariational
thawed and frozen Gaussian approximations, and their special limits based on
the global harmonic, local harmonic, single-Hessian, local cubic, and local
quartic approximations for the potential energy. We also propose a new method
by augmenting the local cubic approximation with a single fourth derivative.
Without substantially increasing the cost, the proposed \textquotedblleft
single-quartic\textquotedblright\ variational Gaussian approximation improves
the accuracy over the local cubic approximation and, at the same time,
conserves both the effective energy and symplectic structure, unlike the much
more expensive local quartic approximation. Most results are presented in both
Heller's\ and Hagedorn's parametrizations of the Gaussian wavepacket.

\end{abstract}
\maketitle

\graphicspath{
{./figures/}{C:/Users/Jiri/Dropbox/Papers/Chemistry_papers/2021/TGA_variational/figures/}}

\section{\label{sec:intro}Introduction}

Semiclassical trajectory-based methods for solving the time-dependent
Schr\"{o}dinger equation (TDSE)\ avoid the exponential scaling of the exact
quantum solution and, in contrast to classical methods, can capture various
quantum effects at least qualitatively. Semiclassical methods have been
successfully applied to the calculation of vibrational and electronic spectra,
fluorescence and internal conversion rates, diffusion constants, and rate
constants of chemical
reactions.\cite{Heller:1981a,book_Mukamel:1999,Miller:2001,book_Tannor:2007,book_Heller:2018}%

Multi-trajectory semiclassical methods, such as the initial value
representation,\cite{Miller:2001} frozen Gaussian
approximation,\cite{Heller:1981} Herman-Kluk
propagator,\cite{Herman_Kluk:1984} phase averaging,\cite{book_Mukamel:1999}
hybrid dynamics,\cite{Grossmann:2006} or
multiple-spawning,\cite{Ben-Nun_Martinez:1998} employ an ensemble of
trajectories and, therefore, account both for the nonlinear spreading and
interference between various parts of the wavepacket. Trajectory ensembles
were even used to capture relativistic effects.\cite{Tsai_Poirier:2016}
Converging such methods numerically, however, often requires many
trajectories, which can become computationally prohibitive if the potential
energy surfaces on which the trajectories evolve are expensive. Unfortunately,
this happens in the most interesting modern applications, where the potential
energy surfaces are evaluated with ab initio electronic structure
codes.\cite{Ben-Nun_Martinez:2000,Ceotto_Atahan:2009,Tatchen_Pollak:2009,Curchod_Martinez:2018}%

Single-trajectory semiclassical approximations, although obviously much
cruder, provide a practical alternative because they avoid the issue of
convergence over the ensemble and permit using accurate potential energy
surfaces. In addition, they provide a simpler physical interpretation and
preserve more geometric properties of the exact solution. Among the earliest
such methods, Heller's \textquotedblleft thawed\textquotedblright\ Gaussian
approximation\cite{Heller:1975,Lee_Heller:1982,Tannor_Heller:1982} propagates
a Gaussian wavepacket along a classical trajectory and lets its width evolve
using the local harmonic approximation for the potential. It is very
efficient, permits an on-the-fly ab initio
implementation,\cite{Wehrle_Vanicek:2014,Wehrle_Vanicek:2015,Begusic_Vanicek:2018a,Begusic_Vanicek:2022}
and includes some anharmonic effects, completely missing in the simpler global
harmonic models.

Although this local harmonic thawed Gaussian wavepacket dynamics (TGWD)
conserves the norm of the wavepacket, this method conserves neither the exact
nor the local harmonic effective energy. Heller appreciated\cite{Heller:1976}
very early the capability of the time-dependent variational
principle\cite{Dirac:1930,book_Frenkel:1934,book_Lubich:2008} to improve the
accuracy of semiclassical dynamics. Applying the variational principle to the
thawed Gaussian ansatz, Coalson and Karplus obtained elegant
equations\cite{Coalson_Karplus:1990} for propagating the Gaussian wavefunction
optimally. Even though this variational Gaussian approximation is
symplectic,\cite{Faou_Lubich:2006,book_Hairer_Wanner:2006,book_Lubich:2008}
conserves the energy exactly, and may even capture shallow tunneling, it does
so at a much higher cost because it requires the expectation values of the
potential energy, gradient, and Hessian. Noting that the local harmonic TGWD
can be also obtained by applying the variational principle to the local
harmonic approximation of the potential, Pattanayak and Schieve improved the
accuracy of Heller's method by including the third derivative of the potential
in their \textquotedblleft extended\textquotedblright%
\ semiclassical\ wavepacket
dynamics.\cite{Pattanayak_Schieve:1994,Pattanayak_Schieve:1994a} Ohsawa and
Leok\cite{Ohsawa_Leok:2013} pointed out that this method is, as the
variational TGWD, but unlike the local harmonic TGWD, exactly symplectic and,
therefore, called it \textquotedblleft symplectic semiclassical wavepacket
dynamics.\textquotedblright%
\cite{Ohsawa_Leok:2013,Ohsawa:2015,Ohsawa:2015a,Ohsawa_Tronci:2017}

To accelerate on-the-fly ab initio applications of the local harmonic TGWD,
Begu\v{s}i\'{c}, Cordova, and Van\'{\i}\v{c}ek
proposed\cite{Begusic_Vanicek:2019} the single-Hessian
approximation,\cite{Begusic_Vanicek:2020,Begusic_Vanicek:2022} in which the
trajectory evolves according to the original potential, but the width of the
wavepacket feels a constant curvature. Remarkably, this simplified
approximation preserves the effective energy exactly and has a Hamiltonian
structure in an augmented phase space.\cite{Begusic_Vanicek:2019}

All of the approximations mentioned in the three preceding paragraphs
propagate a Gaussian wavepacket in a time-dependent quadratic potential whose
parameters depend on the instantaneous state of the system. Therefore, each of
these approximations is also an exact solution of a certain nonlinear TDSE,
i.e., a Schr\"{o}dinger equation whose Hamiltonian depends on the quantum
state. The most famous nonlinear Schr\"{o}dinger equation is probably the
Gross--Pitaevskii equation\cite{Gross:1961,Pitaevskii:1961,Carles:2002}
describing approximately the Bose-Einstein condensates, but many other
approximations can be stated as exact solutions of a TDSE with a
state-dependent Hamiltonian operator.\cite{Begusic_Vanicek:2019} The
time-dependent variational
principle\cite{Dirac:1930,book_Frenkel:1934,book_Lubich:2008} seeks optimal
solutions in a nonlinear manifold of possible solutions and thus yields many
examples of nonlinear TDSEs, including the time-dependent
Hartree,\cite{Dirac:1930,McLachlan:1964,Jungwirth_Gerber:1995} time-dependent
Hartree-Fock,\cite{Dirac:1930,McLachlan_Ball:1964} or multi-configurational
time-dependent Hartree method.\cite{Meyer_Cederbaum:1990, Beck_Jackle:2000}
Most nonlinear TDSEs, however, do not rely on the variational
principle---representative
examples\cite{Roulet_Vanicek:2021,Roulet_Vanicek:2021a} are Heller's thawed
Gaussian approximation\cite{Heller:1975} and the local control
theory,\cite{Kosloff_Tannor:1989,Marquetand_Engel:2007,Engel_Tannor:2009}
which seeks a state-dependent electric field that increases or decreases an
observable of interest.

\begin{table}
\caption{Summary of nonstandard notation.} \label{tab:notation}
\begin{ruledtabular}
\begin{tabular}{lll}
Quantity \& equation, where defined & Symbol \& definition \\
\hline
Shifted position (\ref{eq:x})                & $x              := q-q_{t}$           \\
Shifted position operator (\ref{eq:x_op})          & $\hat{x}        := \hat{q}-q_{t}$     \\
Scaled \& shifted position (\ref{eq:xi})         & $\xi            := A_{t}\cdot x+p_{t}$ \\
Position covariance (\ref{eq:Sigma_t})                & $\Sigma_{t} :=  \operatorname*{Cov}(\hat{q})$ \\
Real part of the width matrix (\ref{eq:Re_A})      & $\mathcal{A} := \operatorname{Re}A$\\
Imag.~part of the width matrix (\ref{eq:Im_A}) & $\mathcal{B} := \operatorname{Im}A$ \\
\end{tabular}
\end{ruledtabular}

\end{table}

This paper explores Gaussian wavepacket dynamics from the perspective of a
nonlinear TDSE with an effective potential that is a quadratic polynomial of
coordinates with state-dependent coefficients. The remainder of the paper is
organized as follows:\ Section~\ref{sec:lin_TDSE} reviews the basic properties
of the linear TDSE in order to highlight the differences from the nonlinear
TDSE, presented in Sec.~\ref{sec:nonlin_TDSE}. The formalism developed in
Sec.~\ref{sec:nonlin_TDSE} is completely general and also applies to nonlinear
TDSEs that do not result from the variational principle and to non-Gaussian
wavepackets. In contrast to the variational or symplectic
approaches,\cite{Faou_Lubich:2006,Ohsawa_Leok:2013} the formalism is
elementary and does not rely on the very elegant, but advanced symplectic
formulation of Hamiltonian quantum dynamics.

The main Sec.~\ref{sec:NL-TDSE_for_GWP} defines the Gaussian wavepacket
dynamics as an example of a nonlinear TDSE. We derive equations of motion for
the parameters of the Gaussian in terms of the coefficients of the effective
quadratic potential and analyze time reversibility and the conservation of
norm, energy, effective energy, and symplectic structure. Although the focus
on Gaussian wavepackets makes it possible to obtain more detailed results, the
formalism remains more general than the variational and symplectic approaches
because the nonlinear TDSE analyzed in Sec.~\ref{sec:NL-TDSE_for_GWP} still
does not have to arise from the variational principle and the equations of
motion do not have to be Hamiltonian for a non-canonical symplectic structure.
Examples of wavepacket dynamics employing Gaussians with a flexible width are
presented in Sec.~\ref{sec:TG_methods} and Gaussians with a fixed width in
Sec.~\ref{sec:FG_methods}. In Sec.~\ref{sec:SQV-TGWD}, we propose a new,
single-quartic TGWD, which---in contrast to the similarly accurate but much
more expensive local quartic approximation---conserves both the symplectic
structure and effective energy and, at the same time, improves the accuracy
over the local cubic approximation without increasing the cost. The proposed
method allows for tunneling, but, unlike Coalson and Karplus's variational
TGWD, makes it possible to evaluate analytically the expectation values of the
potential, gradient, and Hessian; it is, therefore, a natural extension of
Heller's thawed Gaussian approximation, which uses classical trajectories and
cannot describe tunnelling. Geometric integrators for the general Gaussian
wavepacket dynamics are described in Sec.~\ref{sec:integrators}.
Section~\ref{sec:Hagedorn_parametrization} translates all results from
Heller's to Hagedorn's parametrization of the Gaussian wavepacket. Finally,
Sec.~\ref{sec:conclusion} discusses the relationship between three approaches
to the TGWD and concludes the paper. For reference, the nonstandard notation
used in this paper is summarized in Table~\ref{tab:notation}.

\section{\label{sec:lin_TDSE}Linear Schr\"{o}dinger equation}

Let us briefly review the properties of the \emph{linear} time-dependent
Schr\"{o}dinger equation (TDSE)%
\begin{equation}
i\hbar|\dot{\psi}(t)\rangle=\hat{H}|\psi(t)\rangle, \label{eq:tdse}%
\end{equation}
in which the wavepacket $\psi(t)$ is driven by the time-independent Hermitian
linear Hamiltonian operator $\hat{H}$. The Hamiltonian is said to be linear
because it is independent of the state $\psi$ and not because of a linear
dependence on coordinates. Indeed, $\hat{H}$ can be a nonlinear function of
coordinates and still be a linear operator.

The state $\psi(t)$ at time $t$ can be obtained from the initial state
$\psi(0)$ formally as $|\psi(t)\rangle=\hat{U}(t)|\psi(0)\rangle$, where
$\hat{U}(t)=\exp(-it\hat{H}/\hbar)$ is the time evolution operator. Because
$\hat{H}$ is a linear operator, so is $\hat{U}(t)$. The evolution is
time-reversible because $\hat{U}(-t)\hat{U}(t)|\psi(0)\rangle=|\psi
(0)\rangle.$

The exact quantum evolution with a time-independent Hermitian linear
Hamiltonian $\hat{H}$ conserves both the norm $\left\Vert \psi(t)\right\Vert $
of the quantum state and its energy
\begin{equation}
E:=\langle\hat{H}\rangle, \label{eq:E}%
\end{equation}
where $\langle\hat{A}\rangle\equiv\langle\hat{A}\rangle_{\psi(t)}:=\langle
\psi(t)|\hat{A}|\psi(t)\rangle$ denotes the expectation value of operator
$\hat{A}$ in the state $\psi(t)$. Both conservation properties follow from a
general expression%
\begin{equation}
d\langle\hat{A}\rangle/dt=(i\hbar)^{-1}\langle\lbrack\hat{A},\hat{H}%
]\rangle\label{eq:evol_exp_A}%
\end{equation}
for the time dependence of $\langle\hat{A}\rangle$, applied to the identity
operator ($\hat{A}=\hat{1}$) or to the Hamiltonian ($\hat{A}=\hat{H}$). The
linear time evolution also conserves the inner product $\langle\psi
(t)|\phi(t)\rangle$ of two different states.

Here, we will usually assume that the Hamiltonian is separable into a sum
\begin{equation}
\hat{H}=\hat{T}+\hat{V}=T(\hat{p})+V(\hat{q}), \label{eq:ham_op}%
\end{equation}
of a kinetic energy term $\hat{T}\equiv T(\hat{p})$, depending only on
momentum $p$, and potential energy term $\hat{V}\equiv V(\hat{q})$, depending
only on position $q$. Both $q$ and $p$ are $D$-dimensional vectors. We call
Hamiltonians described by Eq.~(\ref{eq:ham_op}) \textquotedblleft
separable,\textquotedblright\ without requiring that the potential energy
$V(q)$ itself be separable into a sum $V_{1}(q_{1})+\cdots+V_{D}(q_{D})$ of
$D$ functions, each depending on a single degree of freedom; beware that many
authors require this property in the definition of separability. While the
potential energy function $V(q)$ can be an arbitrary real-valued function, for
the kinetic energy we shall assume the quadratic form
\begin{equation}
T(p)=p^{T}\cdot m^{-1}\cdot p/2, \label{eq:kin_en}%
\end{equation}
where $m$ is a (not necessarily diagonal) positive-definite real symmetric
$D\times D$ mass matrix. In a linear TDSE, neither $\hat{T}$ nor $\hat{V}$
depends on the state $\psi$.

\section{\label{sec:nonlin_TDSE}Nonlinear Schr\"{o}dinger equation}

\begin{figure}
[htbp]%
\centering\includegraphics[width=\columnwidth]{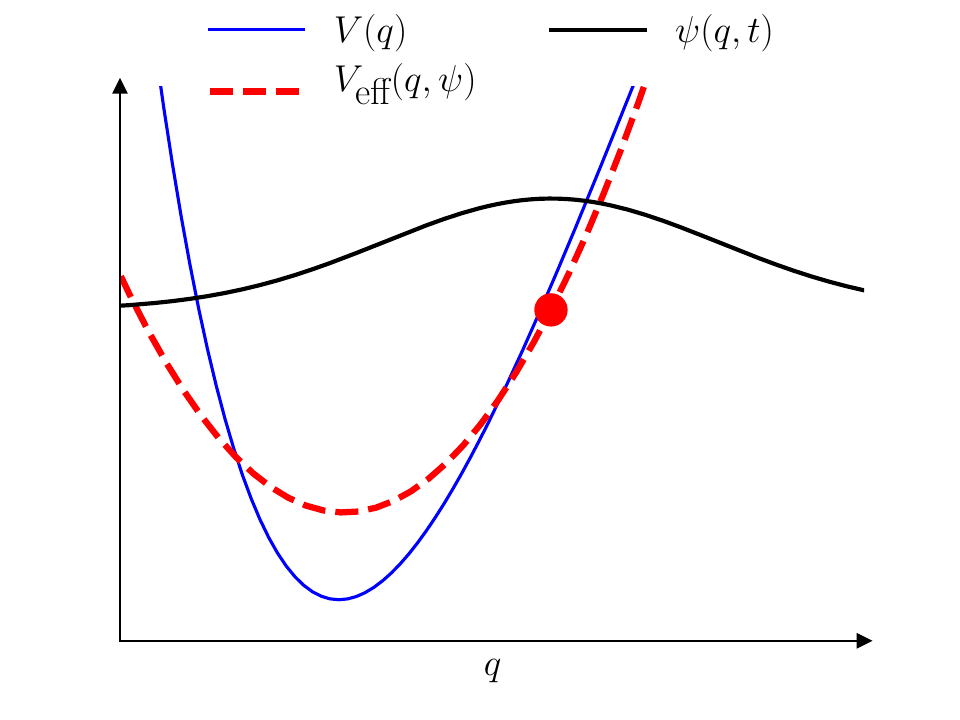}\caption{{Example of a wavepacket $\psi(q,t)$ that solves exactly a nonlinear Schr\"{o}dinger equation
(\ref{eq:nonlinear_tdse}) with a separable, state-dependent effective Hamiltonian $\hat{H}_{\text{eff}}(\psi)=T(\hat
{p})+V_{\text{eff}}(\hat{q};\psi)$. Here, the wavepacket is
a Gaussian (\ref{eq:GWP}) and the effective potential $V_{\text{eff}}(q;\psi)$  is a
quadratic polynomial (\ref{eq:V_eff}) whose coefficients are given by the single quartic variational
approximation [Eqs.~(\ref{eq:V2_SQV-TGWD})-(\ref{eq:V0_SQV-TGWD})] to the original, Morse potential $V(q)$.}}\label{fig:GWP_and_V_eff}%

\end{figure}

When Eq.~(\ref{eq:tdse}) is solved approximately, its approximate solution can
often\cite{book_Lubich:2008,Roulet_Vanicek:2021,Roulet_Vanicek:2021a} be
interpreted as the \emph{exact} solution of a \emph{nonlinear} Schr\"{o}dinger
differential equation%
\begin{equation}
i\hbar|\dot{\psi}(t)\rangle=\hat{H}_{\text{eff}}[\psi(t)]|\psi(t)\rangle
\label{eq:nonlinear_tdse}%
\end{equation}
with an effective Hamiltonian operator $\hat{H}_{\text{eff}}(\psi)$ depending
on the state $\psi$ (see Fig.~\ref{fig:GWP_and_V_eff}). Although one may
envision a more general nonlinearity, where $\hat{H}_{\text{eff}}(\psi
)|\psi\rangle$ in the right-hand side of Eq.~(\ref{eq:nonlinear_tdse}) would
be replaced by an arbitrary functional of $\psi$, the slightly less general
case described by the \textquotedblleft quasi-linear\textquotedblright\ form
of\ Eq.~(\ref{eq:nonlinear_tdse}) is much more interesting because it
preserves some features of the linear Schr\"{o}dinger equation. Notation
$\hat{H}_{\text{eff}}(\psi)|\psi\rangle$ reflects our assumption that whereas
the mapping $\hat{H}_{\text{eff}}:|\psi\rangle\mapsto\hat{H}_{\text{eff}}%
(\psi)|\psi\rangle$ is nonlinear, the mapping $\hat{H}_{\text{eff}}%
(\psi):|\phi\rangle\mapsto\hat{H}_{\text{eff}}(\psi)|\phi\rangle$ is linear
for all $\psi$. In addition, we shall assume that the expectation value
\begin{equation}
\langle\hat{H}_{\text{eff}}(\psi)\rangle_{\phi}:=\langle\phi|\hat
{H}_{\text{eff}}(\psi)\phi\rangle, \label{eq:real_exp_value_Heff}%
\end{equation}
generalizing the energy, is real for any $\psi$ and $\phi$. This condition
implies that $\hat{H}_{\text{eff}}(\psi)$, considered as a linear operator
(i.e., with fixed $\psi$), is Hermitian:%
\begin{equation}
\langle\phi|\hat{H}_{\text{eff}}(\psi)\theta\rangle=\langle\hat{H}%
_{\text{eff}}(\psi)\phi|\theta\rangle\text{ \ \ for all }\psi,\phi,\theta.
\end{equation}

The state at time $t\geq t_{0}$ can be obtained from the initial state at time
$t_{0}$ as $|\psi(t)\rangle=\hat{U}(t,t_{0};\psi)|\psi(t_{0})\rangle$, i.e.,
by the multiplication with the evolution operator%
\begin{equation}
\hat{U}(t,t_{0};\psi)=\mathcal{T}\exp\left[  -\frac{i}{\hbar}\int_{t_{0}}%
^{t}\hat{H}_{\text{eff}}[\psi(t^{\prime})]dt^{\prime}\right]  ,
\end{equation}
where $\mathcal{T}$ indicates time ordering. The evolution operator is
nonlinear because it depends on the propagated state. Yet, the evolution
guided by the nonlinear Schr\"{o}dinger equation (\ref{eq:nonlinear_tdse})
remains time-reversible:
\begin{equation}
\hat{U}(t_{0},t;\psi)\hat{U}(t,t_{0};\psi)|\psi(t_{0})\rangle=|\psi
(t_{0})\rangle, \label{eq:reversibility_nlse}%
\end{equation}
because, for $t_{0}\leq t$,%
\begin{align}
&  \hat{U}(t_{0},t;\psi) =\mathcal{\bar{T}}\exp\left[  -\frac{i}{\hbar}%
\int_{t}^{t_{0}}\hat{H}_{\text{eff}}[\psi(t^{\prime})]dt^{\prime}\right]
\nonumber\\
&  =\mathcal{\bar{T}}\exp\left[  \frac{i}{\hbar}\int_{t_{0}}^{t}\hat
{H}_{\text{eff}}[\psi(t^{\prime})]dt^{\prime}\right]  =\hat{U}(t,t_{0}%
;\psi)^{-1},
\end{align}
where $\mathcal{\bar{T}}$ denotes reverse time ordering. Because the
Hamiltonian is state-dependent, the scalar product generally depends on time:
\begin{align}
\frac{d}{dt}\left\langle \psi|\phi\right\rangle  &  =\langle\dot{\psi}%
|\phi\rangle+\langle\psi|\dot{\phi}\rangle\nonumber\\
&  =(i\hbar)^{-1}\langle\psi|[-\hat{H}_{\text{eff}}(\psi)+\hat{H}_{\text{eff}%
}(\phi)]\phi\rangle\neq0, \label{eq:time_dependence_of_sc_prod}%
\end{align}
where we used relation~(\ref{eq:d_psi_dot_phi_dt}) from
Appendix~\ref{sec:d_psi_dot_phi_dt}\ for $\langle\dot{\psi}|\phi\rangle$.

Let us inspect the conservation of norm and energy by the nonlinear
Schr\"{o}dinger equation~(\ref{eq:nonlinear_tdse}). These and other useful
properties are obtained by generalizing Eq.~(\ref{eq:evol_exp_A}) to
state-dependent operators $\hat{H}_{\text{eff}}(\psi)$ and $\hat{A}(\psi)$. In
Appendix~\ref{sec:time_dependence_of_exp_A}, we prove that the expectation
value $\langle\hat{A}(\psi)\rangle:=\langle\psi|\hat{A}(\psi)\psi\rangle$
evolves in time according to the equation
\begin{equation}
\frac{d\langle\hat{A}(\psi)\rangle}{dt}=\left\langle \frac{d}{dt}\hat{A}%
(\psi)\right\rangle +\frac{1}{i\hbar}\langle\lbrack\hat{A}(\psi),\hat
{H}_{\text{eff}}(\psi)]\rangle. \label{eq:nonlin_evol_exp_nonlin_A}%
\end{equation}
For a linear, state-independent, operator $\hat{A}$, the general
Eq.~(\ref{eq:nonlin_evol_exp_nonlin_A}) simplifies to%
\begin{equation}
d\langle\hat{A}\rangle/dt=(i\hbar)^{-1}\langle\lbrack\hat{A},\hat
{H}_{\text{eff}}(\psi)]\rangle. \label{eq:nonlin_evol_exp_lin_A}%
\end{equation}
Applying this expression, which is analogous to Eq.~(\ref{eq:evol_exp_A}) for
the linear TDSE, to the identity operator demonstrates that the nonlinear TDSE
(\ref{eq:nonlinear_tdse}) conserves the norm:%
\begin{equation}
\frac{d}{dt}\left\Vert \psi\left(  t\right)  \right\Vert ^{2}=\frac{d}%
{dt}\langle\hat{1}\rangle=\frac{1}{i\hbar}\langle\lbrack\hat{1},\hat
{H}_{\text{eff}}(\psi)]\rangle=0, \label{eq:norm_squared_conservation}%
\end{equation}
while its application to the exact Hamiltonian yields the time dependence of
energy:%
\begin{equation}
\dot{E}=d\langle\hat{H}\rangle/dt=(i\hbar)^{-1}\langle\lbrack\hat{H},\hat
{H}_{\text{eff}}(\psi)]\rangle. \label{eq:dE_dt}%
\end{equation}
As a result, the energy $E$ may not be conserved under the evolution with the
effective Hamiltonian $\hat{H}_{\text{eff}}(\psi)$.

For effective Hamiltonians, one can also study the time dependence of the
effective energy%
\begin{equation}
E_{\text{eff}}:=\langle\hat{H}_{\text{eff}}(\psi)\rangle. \label{eq:E_eff}%
\end{equation}
Because the exact energy (\ref{eq:E})\ is conserved under the exact time
evolution with Hamiltonian $\hat{H}$, one might expect that the effective
energy (\ref{eq:E_eff}) would be conserved under the evolution with the
effective Hamiltonian $\hat{H}_{\text{eff}}$. This is not true in general;
taking $\hat{A}=\hat{H}_{\text{eff}}(\psi)$ in
Eq.~(\ref{eq:nonlin_evol_exp_nonlin_A}) gives%
\begin{equation}
\dot{E}_{\text{eff}}=\langle d\hat{H}_{\text{eff}}[\psi(t)]/dt\rangle.
\label{eq:dEeff_dt1}%
\end{equation}

\subsection{\label{subsec:nonlin_TDSE_w_separ_H}Nonlinear TDSE with a
separable Hamiltonian}

In what follows, we only consider separable Hamiltonians (\ref{eq:ham_op}) and
related separable effective Hamiltonians
\begin{equation}
\hat{H}_{\text{eff}}(\psi)=\hat{T}+\hat{V}_{\text{eff}}(\psi)=T(\hat
{p})+V_{\text{eff}}(\hat{q};\psi). \label{eq:H_eff}%
\end{equation}
Expressed in position representation, the nonlinear Schr\"{o}dinger
equation~(\ref{eq:nonlinear_tdse}) becomes%
\begin{equation}
i\hbar\partial_{t}\psi(q,t)=-\frac{\hbar^{2}}{2}\nabla^{T}\cdot m^{-1}%
\cdot\nabla\psi(q,t)+V_{\text{eff}}(q;\psi)\psi\left(  q,t\right)  .
\label{eq:tdse_q-rep}%
\end{equation}
With the obvious exception of energy conservation, many following results can
be easily generalized to explicitly time-dependent potentials $\hat{V}\left(
t\right)  $. However, for the sake of brevity, we shall continue assuming that
the original potential energy $\hat{V}$ is independent of time and that the
effective potential $\hat{V}_{\text{eff}}\equiv\hat{V}_{\text{eff}}(\psi)$
depends on time only implicitly, via the dependence on the state $\psi(t)$.

For separable Hamiltonians, general expressions~(\ref{eq:dE_dt}) and
(\ref{eq:dEeff_dt1}) for the time dependence of energy and effective energy
reduce to (see Appendix~\ref{sec:time_dependence_of_energy_in_separ_H} for
proof)%
\begin{align}
\dot{E}  &  =\operatorname{Re}\langle\hat{p}^{T}\cdot m^{-1}\cdot(\hat
{V}^{\prime}-\hat{V}_{\text{eff}}^{\prime})\rangle\text{ \ and}%
\label{eq:comm_H_Heff_1}\\
\dot{E}_{\text{eff}}  &  =\langle d\hat{V}_{\text{eff}}[\psi(t)]/dt\rangle,
\label{eq:dEeff_dt}%
\end{align}
where the gradient of the effective potential is defined as%
\begin{equation}
\hat{V}_{\text{eff}}^{\prime}:=\left.  \partial V_{\text{eff}}(q;\psi
)/\partial q\right\vert _{q=\hat{q}}.
\end{equation}

Finally, applying Eq.~(\ref{eq:nonlin_evol_exp_lin_A}) to the position and
momentum operators shows that the Ehrenfest theorem continues to hold for the
nonlinear Schr\"{o}dinger equation with the separable Hamiltonian
(\ref{eq:H_eff}): namely,%
\begin{align}
d\langle\hat{q}\rangle/dt  &  =m^{-1}\cdot\langle\hat{p}\rangle
,\label{eq:nonlinear_Ehrenfest_q}\\
d\langle\hat{p}\rangle/dt  &  =-\langle\hat{V}_{\text{eff}}^{\prime}\rangle.
\label{eq:nonlinear_Ehrenfest_p}%
\end{align}
These two equations follow from the explicit expressions
(\ref{eq:comm_q_and_H_eff}) and (\ref{eq:comm_p_and_H_eff}) in
Appendix~\ref{sec:time_dependence_of_energy_in_separ_H}\ for the commutators
of position and momentum with the effective Hamiltonian.

\section{\label{sec:NL-TDSE_for_GWP}Nonlinear time-dependent Schr\"{o}dinger
equation for a Gaussian wavepacket}

\subsection{\label{subsec:gaussian}Gaussian wavepacket}

Let us consider approximate solutions of the linear Schr\"{o}dinger
Eq.~(\ref{eq:tdse}), which are exact solutions of the nonlinear
Schr\"{o}dinger Eq.~(\ref{eq:nonlinear_tdse}) and in which the wavepacket
$\psi(t)$ has a Gaussian form at all times. I.e., we will consider wavepackets
written in position representation as
\begin{equation}
\psi(q,t)=\exp\left[  \frac{i}{\hbar}\left(  \frac{1}{2}x^{T}\cdot A_{t}\cdot
x+p_{t}^{T}\cdot x+\gamma_{t}\right)  \right]  , \label{eq:GWP}%
\end{equation}
where the shifted position vector
\begin{equation}
x:=q-q_{t} \label{eq:x}%
\end{equation}
was introduced to simplify notation and $q_{t}$, $p_{t}$, $A_{t}$, and
$\gamma_{t}$ are Heller's\cite{Heller:1975,Heller:1976a,Heller:2006}
parameters determining the time dependence of $\psi(t)$. Parameters $q_{t}$
and $p_{t}$ are real $D$-dimensional vectors equal to the expectation values
of position and momentum, $A_{t}$ is a complex symmetric $D\times D$ matrix,
and $\gamma_{t}$ is a complex scalar. The positive definite imaginary part of
$A_{t}$ determines the width of the Gaussian, while its real part introduces a
spatial chirp; the real part of $\gamma_{t}$ gives a time-dependent phase,
while its imaginary part controls the norm of $\psi(t)$, given by%
\begin{equation}
\left\Vert \psi(t)\right\Vert =\det\left(  \operatorname{Im}A_{t}/\pi
\hbar\right)  ^{-1/4}e^{-\operatorname{Im}\gamma_{t}/\hbar}.
\label{eq:norm_squared_GWP}%
\end{equation}
As shown above, in Eq.~(\ref{eq:norm_squared_conservation}), this norm is
conserved by the solutions of Eq.~(\ref{eq:nonlinear_tdse}). Choosing
$\gamma_{0}$ so that
\begin{equation}
\exp\left(  -\operatorname{Im}\gamma_{0}/\hbar\right)  =[\det
(\operatorname{Im}A_{0}/\pi\hbar)]^{1/4} \label{eq:initial_GWP_normalization}%
\end{equation}
ensures unit normalization at all times.

\subsection{\label{subsec:nonlin_TDSE_gauss}Nonlinear time-dependent
Schr\"{o}dinger equation in terms of parameters of the Gaussian wavepacket}

Now we shall show that the nonlinear Schr\"{o}dinger equation
(\ref{eq:tdse_q-rep}) is solved exactly by a Gaussian wavepacket
(\ref{eq:GWP}) if an only if the effective potential energy operator $\hat
{V}_{\text{eff}}(\psi)$ is a quadratic polynomial
\begin{align}
\hat{V}_{\text{eff}}(\psi)  &  \equiv V_{\text{eff}}(\hat{q};\psi)\equiv
V_{\text{eff}}(\hat{x};\psi)\nonumber\\
&  =V_{0}+V_{1}^{T}\cdot\hat{x}+\hat{x}^{T}\cdot V_{2}\cdot\hat{x}%
/2\nonumber\\
&  =V_{0}+V_{1}^{T}\cdot\hat{x}+\operatorname{Tr}(V_{2}\cdot\hat{x}\otimes
\hat{x}^{T})/2 \label{eq:V_eff}%
\end{align}
in the shifted position operator
\begin{equation}
\hat{x}:=\hat{q}-q_{t} \label{eq:x_op}%
\end{equation}
with coefficients $V_{j}\equiv V_{j}(\psi)$ \ ($j=0,1,2$) that may depend on
the state $\psi$. Note that $\hat{V}_{\text{eff}}(\psi)$ depends on $\psi$ not
only through the coefficients $V_{j}(\psi)$ but also through $\hat{x}=\hat
{q}-q_{t}$ since $q_{t}=\left\langle \hat{q}\right\rangle _{\psi}$. See
Fig.~\ref{fig:GWP_and_V_eff} for an example.

Let us rewrite the Schr\"{o}dinger equation (\ref{eq:tdse_q-rep}) in terms of
ordinary differential equations for the parameters of the Gaussian
(\ref{eq:GWP}). Using the chain rule and Eqs.~(\ref{eq:der_psi_wrt_x}) and
(\ref{eq:der_psi_wrt_A}) from Appendix~\ref{sec:properties_of_GWP}, the time
derivative of the Gaussian wavepacket~(\ref{eq:GWP}) is%
\begin{align}
\partial\psi/\partial t  &  =\left(  \partial\psi/\partial q_{t}\right)
^{T}\cdot\dot{q}_{t}+\left(  \partial\psi/\partial p_{t}\right)  ^{T}\cdot
\dot{p}_{t}\nonumber\\
&  +\operatorname*{Tr}[\left(  \partial\psi/\partial A_{t}\right)  ^{T}%
\cdot\dot{A}_{t}]+(\partial\psi/\partial\gamma_{t})\dot{\gamma}_{t}.
\label{eq:dpsi_dt}%
\end{align}
Employing expressions (\ref{eq:dpsi_dq})-(\ref{eq:dpsi_dgamma}) for the
partial derivatives of $\psi$ with respect to various parameters, we can
rewrite the time derivative (\ref{eq:dpsi_dt}) as
\begin{equation}
i\hbar\frac{\partial\psi}{\partial t}=\left(  \xi^{T}\cdot\dot{q}_{t}%
-x^{T}\cdot\dot{p}_{t}-\frac{1}{2}x^{T}\cdot\dot{A}_{t}\cdot x+\dot{\gamma
}_{t}\right)  \psi, \label{eq:dpsi_dt_in_q_rep}%
\end{equation}
where we introduced a complex vector
\begin{equation}
\xi:=A_{t}\cdot x+p_{t}=A_{t}\cdot(q-q_{t})+p_{t} \label{eq:xi}%
\end{equation}
to simplify notation. The kinetic energy acting on $\psi$ requires
differentiating $\psi$ twice with respect to $x$:%
\begin{align}
&  \langle q|\hat{T}|\psi\rangle=(-\hbar^{2}/2)\,\nabla^{T}\cdot m^{-1}%
\cdot\nabla\psi(q)\nonumber\\
&  =\frac{1}{2}\left[  \xi^{T}\cdot m^{-1}\cdot\xi-i\hbar\operatorname*{Tr}%
\left(  m^{-1}\cdot A_{t}\right)  \right]  \psi(q), \label{eq:q_T_psi}%
\end{align}
where we used Eq.~(\ref{eq:2T_psi}) from Appendix~\ref{sec:properties_of_GWP}.

Schr\"{o}dinger equation~(\ref{eq:nonlinear_tdse})\ can be also written as%
\begin{equation}
0=i\hbar|\dot{\psi}(t)\rangle-\hat{H}_{\text{eff}}|\psi(t)\rangle\text{
\ \ and,} \label{eq:tdse_zero}%
\end{equation}
using Eqs.~(\ref{eq:dpsi_dt_in_q_rep}) and (\ref{eq:q_T_psi}), in position
representation as%
\begin{equation}
0=\left[  f(x)-V_{\text{eff}}(x)\right]  \psi, \label{eq:tdse_f-V}%
\end{equation}
where%
\begin{equation}
f(x)=C_{0}+C_{1}^{T}\cdot x+x^{T}\cdot C_{2}\cdot x/2 \label{eq:quad_pol_f}%
\end{equation}
is a quadratic polynomial with coefficients%
\begin{align}
C_{0}(\psi)  &  :=p_{t}^{T}\cdot\dot{q}_{t}-\dot{\gamma}_{t}+(i\hbar
/2)\operatorname*{Tr}\left(  m^{-1}\cdot A_{t}\right) \nonumber\\
&  ~~~-p_{t}^{T}\cdot m^{-1}\cdot p_{t}/2,\label{eq:C0}\\
C_{1}(\psi)  &  :=A_{t}\cdot\dot{q}_{t}-\dot{p}_{t}-A_{t}\cdot m^{-1}\cdot
p_{t},\label{eq:C1}\\
C_{2}(\psi)  &  :=-\dot{A}_{t}-A_{t}\cdot m^{-1}\cdot A_{t}. \label{eq:C2}%
\end{align}
Because $f(x)$ is a quadratic polynomial, Eq.~(\ref{eq:tdse_f-V}) is satisfied
at all $x$ if and only if $V_{\text{eff}}(x;\psi)$ is also a quadratic
polynomial in $x$ in the form of Eq.~(\ref{eq:V_eff}) and, in addition,
$C_{j}(\psi)=V_{j}(\psi)$ for $j=0,1,2$. Let us summarize this in

\textbf{Proposition 1 (Gaussian wavepacket in a linear or nonlinear TDSE)}.
Gaussian wavepacket (\ref{eq:GWP}) solves the nonlinear TDSE
(\ref{eq:tdse_q-rep}) with a possibly state-dependent effective potential
$\hat{V}_{\text{eff}}(\psi)$ if and only if $V_{\text{eff}}$ is a quadratic
potential (\ref{eq:V_eff}) and $C_{j}(\psi)=V_{j}(\psi)$ for $j=0,1,2$. In
particular, the Gaussian wavepacket solves the linear TDSE (\ref{eq:tdse}) if
and only if the linear (i.e., independent of $\psi$) operator $\hat{V}$ is a
quadratic polynomial of coordinates.

The system of equations $C_{j}(\psi)=V_{j}(\psi)$ for $j=0,1,2$ seems rather
complicated to be useful in practice because Eqs.~(\ref{eq:C0})-(\ref{eq:C2})
couple the time derivatives of the Gaussian parameters. However, it is easy to
invert this system:

\textbf{Proposition 2.} Let $V_{0}(\psi)$, $V_{1}(\psi)$, and $V_{2}(\psi)$
be, respectively, some prescribed real scalar, vector, and symmetric matrix
functions of the state $\psi$. Then the Gaussian wavepacket (\ref{eq:GWP})
solves the nonlinear TDSE (\ref{eq:tdse_q-rep}) with the effective potential
(\ref{eq:V_eff}) if and only if the parameters of the Gaussian solve the
following system of ordinary differential equations:
\begin{align}
\dot{q}_{t}  &  =m^{-1}\cdot p_{t},\label{eq:q_dot}\\
\dot{p}_{t}  &  =-V_{1},\label{eq:p_dot}\\
\dot{A}_{t}  &  =-A_{t}\cdot m^{-1}\cdot A_{t}-V_{2}\,,\label{eq:A_dot}\\
\dot{\gamma}_{t}  &  =T(p_{t})-V_{0}+(i\hbar/2)\operatorname*{Tr}\left(
m^{-1}\cdot A_{t}\right)  . \label{eq:gamma_dot}%
\end{align}

\emph{Proof.} Proposition 1 implies that we can replace $C_{j}$ with $V_{j}$,
$j=0,1,2$ in Eqs. (\ref{eq:C0})-(\ref{eq:C2}). Equation (\ref{eq:A_dot}) of
motion for $\dot{A}_{t}$ follows immediately by inverting Eq.~(\ref{eq:C2})
for the $C_{2}$ matrix. Equation (\ref{eq:q_dot}) for $\dot{q}_{t}$ is
obtained from the imaginary part of Eq.~(\ref{eq:C1}) for the $C_{1}$ vector:%
\begin{equation}
0=\operatorname{Im}A_{t}\cdot\dot{q}_{t}-\operatorname{Im}A_{t}\cdot
m^{-1}\cdot p_{t}. \label{eq:Im_C1}%
\end{equation}
Because $\psi(t)$ conserves the norm (\ref{eq:norm_squared_GWP}) and the
initial state $\psi(0)$ is normalized, $\operatorname{Im}A_{t}$ must be
invertible. Multiplying the last equation on the left with $\left(
\operatorname{Im}A_{t}\right)  ^{-1}$ yields Eq.~(\ref{eq:q_dot}) for $\dot
{q}_{t}$. Substituting the equation of motion (\ref{eq:q_dot}) for $\dot
{q}_{t}$ into Eq.~(\ref{eq:C1}) for $C_{1}$ and Eq.~(\ref{eq:C0}) for $C_{0}$
yields Eq.~(\ref{eq:p_dot}) for $\dot{p}_{t}$ and Eq.~(\ref{eq:gamma_dot}) for
$\dot{\gamma}_{t}$. The opposite implication is proved similarly: In
particular, Eq.~(\ref{eq:C0}) follows from Eqs.~(\ref{eq:q_dot}) and
(\ref{eq:gamma_dot}), whereas Eqs.~(\ref{eq:q_dot}) and (\ref{eq:p_dot}) imply
Eq.~(\ref{eq:C1}) and Eq.~(\ref{eq:A_dot}) implies Eq.~(\ref{eq:C2}).

Next, after discussing various properties of the effective
potential~(\ref{eq:V_eff}), we will show that Eqs.~(\ref{eq:q_dot}) and
(\ref{eq:p_dot}) of motion for position and momentum also follow, more
generally and directly, from the Ehrenfest theorem.

\subsection{Properties of the quadratic effective potential}

Let us list several useful properties of the effective potential
$V_{\text{eff}}$ of Eq.~(\ref{eq:V_eff}). Since $x:=q-q_{t}$, the gradient
vector and Hessian matrix of $V_{\text{eff}}$ are%
\begin{align}
V_{\text{eff}}^{\prime}(x)  &  =V_{1}+V_{2}\cdot x,\label{eq:V_prime_eff}\\
V_{\text{eff}}^{\prime\prime}(x)  &  =V_{2}. \label{eq:V_prime_prime_eff}%
\end{align}
Here, we have used and will use short-hand notations%
\begin{align}
g^{\prime}  &  :=\operatorname{grad}g:=\nabla g,\\
g^{\prime\prime}  &  :=\operatorname*{Hess}g:=\nabla\otimes\nabla^{T}g,
\end{align}
for the gradient and Hessian of function $g(x)$. Expected values of the
effective potential energy (\ref{eq:V_eff}), its gradient, and its Hessian in
the Gaussian wavepacket~(\ref{eq:GWP}) are%
\begin{align}
\langle\hat{V}_{\text{eff}}\rangle &  =V_{0}+V_{1}^{T}\cdot\langle\hat
{x}\rangle+\operatorname{Tr}\left[  V_{2}\cdot\langle\hat{x}\otimes\hat{x}%
^{T}\rangle\right]  /2\nonumber\\
&  =V_{0}+\operatorname{Tr}(V_{2}\cdot\Sigma_{t})/2,\label{eq:exp_value_V_eff}%
\\
\langle\hat{V}_{\text{eff}}^{\prime}\rangle &  =V_{1}+V_{2}\cdot\langle\hat
{x}\rangle=V_{1},\label{eq:exp_value_grad_V_eff}\\
\langle\hat{V}_{\text{eff}}^{\prime\prime}\rangle &  =V_{2},
\label{eq:exp_value_Hess_V_eff}%
\end{align}
where we have invoked relations
\begin{align}
\langle\hat{x}\rangle &  =\langle\hat{q}-q_{t}\rangle=\langle\hat{q}%
\rangle-q_{t}=0,\\
\langle\hat{x}\otimes\hat{x}^{T}\rangle &  =\langle(\hat{q}-q_{t})\otimes
(\hat{q}-q_{t})^{T}\rangle=\operatorname{Cov}(\hat{q})
\end{align}
for the mean and covariance of position and introduced a short-hand notation%
\begin{equation}
\Sigma_{t}:=\operatorname{Cov}(\hat{q}). \label{eq:Sigma_t}%
\end{equation}

In a Gaussian wavepacket, $q_{t}=\langle\hat{q}\rangle$, $p_{t}=\langle\hat
{p}\rangle$, and therefore the equations of motion~(\ref{eq:q_dot}) and
(\ref{eq:p_dot}) for position and momentum also follow immediately from
Eqs.~(\ref{eq:nonlinear_Ehrenfest_q}) and (\ref{eq:nonlinear_Ehrenfest_p}) of
the Ehrenfest theorem for the general nonlinear TDSE (\ref{eq:tdse_q-rep}) and
from Eq.~(\ref{eq:exp_value_grad_V_eff}) for $\langle\hat{V}_{\text{eff}%
}^{\prime}\rangle$. Ehrenfest theorem was also used by Pattanayak and Schieve
to derive the equations of semiquantal
dynamics.\cite{Pattanayak_Schieve:1994,Pattanayak_Schieve:1994a} Finally, note
that for a fixed state $\psi$, $\hat{V}_{\text{eff}}(\psi)$ is a Hermitian
operator because
\begin{equation}
\langle\hat{V}_{\text{eff}}(\psi)\rangle_{\phi}=V_{0}(\psi)+\operatorname{Tr}%
\left[  V_{2}(\psi)\cdot\operatorname{Cov}_{\phi}(\hat{q})\right]  /2
\end{equation}
is real for any $\psi$ and $\phi$ since $V_{0}(\psi)$, $V_{2}(\psi)$, and the
covariance $\operatorname{Cov}_{\phi}(\hat{q})$ of position in the state
$\phi$ are all real. This justifies our assumption
(\ref{eq:real_exp_value_Heff}) of hermiticity of $\hat{H}_{\text{eff}}$ used
in the general analysis in Sec.~\ref{sec:nonlin_TDSE}.

\section{\label{sec:geom_properties}Geometric properties of Gaussian
wavepacket dynamics}

As mentioned in Sec.~\ref{sec:lin_TDSE}, the exact solution of the linear TDSE
(\ref{eq:tdse}) with a time-independent Hamiltonian has several
\textquotedblleft geometric\textquotedblright\ properties: the time evolution
is linear, unitary, norm-conserving, energy-conserving, symplectic, and
time-reversible. Symplecticity means that the time evolution conserves the
symplectic structure---a symplectic $2$-form $\omega(\psi,\phi):=-2\hbar
\operatorname{Im}\langle\psi,\phi\rangle$, defined on the Hilbert space as the
imaginary part of the scalar product.\cite{book_Lubich:2008} The loss of
linearity implies that the nonlinear TDSE fails to conserve the inner product.
As a result, conservation of neither the norm nor symplectic structure is
guaranteed. Let us discuss the time reversibility and conservation of norm,
energy, effective energy, and symplectic structure by the Gaussian wavepacket
dynamics---the nonlinear TDSE (\ref{eq:tdse_q-rep}) with the effective
potential (\ref{eq:V_eff}).

\subsection{Norm conservation}

As already shown in Secs.~\ref{sec:nonlin_TDSE}\ and \ref{subsec:gaussian},
the norm of a Gaussian wavepacket is always conserved although a scalar
product between two different initial states is not.

\subsection{Exact and effective energies of a Gaussian wavepacket}

Quantum-mechanical energy in a state $\psi$ driven by the separable
Hamiltonian (\ref{eq:ham_op}) is given by the sum of expectation values of
kinetic and potential energies:%
\begin{equation}
E=\langle\hat{H}\rangle=\langle\hat{T}\rangle+\langle\hat{V}\rangle.
\label{eq:exp_H}%
\end{equation}
In general, the expected value $\langle\hat{V}\rangle$ of the potential energy
cannot be simplified. Because the kinetic energy has the quadratic form
(\ref{eq:kin_en}), its expected value in a Gaussian is%
\begin{align}
\langle\hat{T}\rangle &  =\left\langle \hat{p}^{T}\cdot m^{-1}\cdot\hat
{p}\right\rangle /2=\operatorname{Tr}\left(  m^{-1}\cdot\left\langle \hat
{p}\otimes\hat{p}^{T}\right\rangle \right)  /2\nonumber\\
&  =T(p_{t})+\operatorname{Tr}[m^{-1}\cdot\operatorname{Cov}(\hat{p})]/2,
\label{eq:exp_T}%
\end{align}
where the first term is the classical kinetic energy at the wavepacket's
center and the second term reflects the finite width of the wavepacket;
$\operatorname{Cov}(\hat{p})$ is the momentum covariance
matrix~(\ref{eq:Cov_p}). In the third step of the derivation, we used
Eq.~(\ref{eq:exp_p_pT}) from Appendix~\ref{sec:covariances}.

Unlike the energy, the effective energy (\ref{eq:E_eff}),%
\begin{equation}
E_{\text{eff}}=\langle\hat{H}_{\text{eff}}\rangle=\langle\hat{T}%
\rangle+\langle\hat{V}_{\text{eff}}\rangle, \label{eq:exp_H_eff}%
\end{equation}
can be evaluated fully analytically since, for quadratic effective potentials
(\ref{eq:V_eff}), $\langle\hat{V}_{\text{eff}}\rangle$ is given by
Eq.~(\ref{eq:exp_value_V_eff}).

Because the effective potential is different from the exact potential,
generally $\langle\hat{V}_{\text{eff}}\rangle\neq\langle\hat{V}\rangle$ and,
therefore, $E_{\text{eff}}\neq E$. Below, we shall see that in the special
case of the variational Gaussian wavepacket dynamics, a beautiful cancellation
results in the equality $E_{\text{eff}}=E$.

\subsection{\label{subsec:time_dependence_of_E}Time dependence or conservation
of energy}

As follows from a more universal analysis in
Sec.~\ref{subsec:nonlin_TDSE_w_separ_H}, evolution of a Gaussian wavepacket
with approximate effective Hamiltonian (\ref{eq:H_eff}) may not conserve
energy. It is a remarkable fact that energy \emph{is} conserved in the special
cases of the variational thawed and frozen Gaussian wavepacket dynamics,
discussed below, in Secs.~\ref{sec:V_TGWD} and \ref{sec:V_FGWD}. More
generally, energy is conserved along the solutions satisfying the
Dirac-Frenkel variational principle for any, not necessarily Gaussian,
wavefunction ansatz compatible with the principle (see
Appendix~\ref{sec:cons_E_norm_by_DFVP}%
).\cite{book_Lubich:2008,Lasser_Lubich:2020,Hackl_Cirac:2020,Lasser_Su:2021}
No other example of Gaussian wavepacket dynamics among those presented in
Secs.~\ref{sec:TG_methods} and \ref{sec:FG_methods} conserves energy.

To see when energy may be conserved, let us derive a universal expression for
the time dependence of energy of a system propagated in a general quadratic
effective potential (\ref{eq:V_eff}). Substituting the gradient
(\ref{eq:V_prime_eff}) of the effective potential into
Eq.~(\ref{eq:comm_H_Heff_1}) gives%
\begin{align}
\dot{E}  &  =\operatorname{Re}\langle\hat{p}^{T}\cdot m^{-1}\cdot(\hat
{V}^{\prime}-V_{1}-V_{2}\cdot\hat{x})\rangle\nonumber\\
&  =\operatorname{Tr}[m^{-1}\cdot\operatorname{Re}\langle(\hat{V}^{\prime
}-V_{1}-V_{2}\cdot\hat{x})\otimes\hat{p}^{T}\rangle]. \label{eq:dE_dt_2}%
\end{align}
Substitution of expression~(\ref{eq:dE_dt_auxiliary}) from
Appendix~\ref{sec:E_dot_in_TGWD} for the expected value in
Eq.~(\ref{eq:dE_dt_2}) yields%
\begin{align}
\dot{E}  &  =p_{t}^{T}\cdot m^{-1}\cdot(\langle\hat{V}^{\prime}\rangle
-V_{1})\nonumber\\
&  ~~~+\operatorname{Tr}[m^{-1}\cdot(\langle\hat{V}^{\prime\prime}%
\rangle-V_{2})\cdot\operatorname{Cov}_{R}(\hat{q},\hat{p})],
\label{eq:dE_dt_3}%
\end{align}
where $\operatorname{Cov}_{R}(\hat{q},\hat{p})=(\hbar/2)\left(
\operatorname{Im}A_{t}\right)  ^{-1}\cdot\operatorname{Re}A_{t}$ is the real
covariance~(\ref{eq:CovR_qp}). Relation (\ref{eq:dE_dt_3}) helps determining
when the energy is conserved exactly. First consider a Gaussian with a purely
imaginary width matrix $A_{t}$. Then $\operatorname{Cov}_{R}(\hat{q},\hat
{p})=0$ and, in order that $\dot{E}$ be zero for arbitrary $m$ and $p_{t}$, we
must, in addition, have $V_{1}=\langle\hat{V}^{\prime}\rangle$. If $A_{t}$ is
purely imaginary, Eq.~(\ref{eq:A_dot}) implies that both $\operatorname{Re}%
A_{t}$ and $\operatorname{Cov}_{R}(\hat{q},\hat{p})$ will become nonzero
unless $\dot{A}_{t}=0$, i.e., we have a \textquotedblleft
frozen\textquotedblright\ Gaussian with constant width matrix $A_{t}=A_{0}$
and require that $V_{2}=-A_{0}\cdot m^{-1}\cdot A_{0}$. If $A_{t}$ is not
purely imaginary, $\operatorname{Cov}_{R}(\hat{q},\hat{p})\neq0$ and we must,
in addition, have $V_{2}=\langle\hat{V}^{\prime\prime}\rangle.$ To sum up,
there are two general ways to guarantee the conservation of energy: Either%
\begin{equation}
V_{1}=\langle\hat{V}^{\prime}\rangle\text{ \ \ and \ \ }V_{2}=\langle\hat
{V}^{\prime\prime}\rangle
\end{equation}
for a Gaussian wavepacket with a flexible width, or%
\begin{equation}
V_{1}=\langle\hat{V}^{\prime}\rangle\text{, }V_{2}=-A_{0}\cdot m^{-1}\cdot
A_{0}\text{, and }\operatorname{Re}A_{0}=0
\end{equation}
for a Gaussian wavepacket with a fixed width. As we shall see, these two cases
occur, respectively, in the variational thawed and frozen Gaussian wavepacket dynamics.

\subsection{Time dependence of the effective energy}

In Sec.~\ref{subsec:nonlin_TDSE_w_separ_H}, we have also seen that the
effective energy of a nonlinear TDSE is not always conserved. To find the time
dependence of the effective energy (\ref{eq:exp_H_eff}) for the effective
potential (\ref{eq:V_eff}), we need the time derivative
\begin{equation}
d\hat{V}_{\text{eff}}/dt=\dot{V}_{0}+\dot{V}_{1}^{T}\cdot\hat{x}-V_{1}%
^{T}\cdot\dot{q}_{t}-\dot{q}_{t}^{T}\cdot V_{2}\cdot\hat{x}+\hat{x}^{T}%
\cdot\dot{V}_{2}\cdot\hat{x}/2. \label{eq:dV_eff_dt}%
\end{equation}
Because the second and fourth terms of Eq.~(\ref{eq:dV_eff_dt}) vanish under
the expectation value, substitution of Eq.~(\ref{eq:dV_eff_dt}) into
Eq.~(\ref{eq:dEeff_dt}) for the time derivative of $E_{\text{eff}}$ yields%
\begin{equation}
\dot{E}_{\text{eff}}=\dot{V}_{0}-V_{1}^{T}\cdot\dot{q}_{t}+\operatorname{Tr}%
(\dot{V}_{2}\cdot\Sigma_{t})/2. \label{eq:d_exp_Heff_dt}%
\end{equation}
The effective energy of the Gaussian generally depends on time. Yet, we will
see that in many examples of the Gaussian wavepacket dynamics, the effective
energy is conserved due to the cancellation of various terms in
Eq.~(\ref{eq:d_exp_Heff_dt}). The effective energy is conserved, e.g., if
Eqs.~(\ref{eq:q_dot})-(\ref{eq:gamma_dot}) of motion for $q_{t},p_{t}$,
$A_{t}$, and $\gamma_{t}$ coincide with Hamilton's equations for the
Hamiltonian $E_{\text{eff}}$ on a symplectic manifold of Gaussian
wavepackets.\cite{Ohsawa_Leok:2013,Vanicek:symplecticity}

\subsection{Time reversibility}

If we denote by $\Lambda:=(q,p,A,\gamma)$ the collection of parameters of the
Gaussian $\psi$, the time evolution of $\psi(t)$ can be expressed in terms of
the time evolution $\Lambda_{t}=\Phi(\Lambda_{0},t)$ of the parameters by a
flow $\Phi(\Lambda,t)$.\ Time reversibility (\ref{eq:reversibility_nlse}) of
the TGWD, equivalent to the condition%
\begin{equation}
\Lambda_{0}=\Phi(\Lambda_{t},-t), \label{eq:reversibility_TGWD}%
\end{equation}
follows from the reversibility (\ref{eq:reversibility_nlse}) of general
nonlinear TDSE. In Sec.~\ref{sec:integrators}, we will provide a more explicit
proof based on condition (\ref{eq:reversibility_TGWD}) and the fact that
symmetric composition of reversible flows is reversible.

\subsection{Symplecticity}

The family of Gaussian wavepackets~(\ref{eq:GWP}) parametrized by $q_{t}$,
$p_{t}$, $A_{t}$, and $\gamma_{t}$ forms a finite-dimensional symplectic
submanifold of the Hilbert space and is equipped with a certain noncanonical
symplectic structure.\cite{Ohsawa_Leok:2013} Ohsawa and Leok showed that
symplectic reduction associated to norm conservation leads to a simpler
symplectic form%
\begin{equation}
\omega=\sum_{j}dq_{j}\wedge dp_{j}+(\hbar/4)\sum_{j,k}d(\mathcal{B}^{-1}%
)_{jk}\wedge d\mathcal{A}_{kj} \label{eq:omega_Heller_reduced}%
\end{equation}
on a manifold with coordinates $\Lambda:=(q,p,\mathcal{A}\equiv
\operatorname{Re}A,\mathcal{B}\equiv\operatorname{Im}A)$; this symplectic
structure is conserved, e.g., by the variational Gaussian approximation, but
not by the original thawed Gaussian approximation.\cite{Ohsawa_Leok:2013} If
the effective Hamiltonian is defined as the expectation value of the exact or
some approximate Hamiltonian, i.e., $H_{\text{eff}}:=\langle\hat{H}%
\rangle_{\psi}$ or $H_{\text{eff}}:=\langle\hat{H}_{\text{appr}}\rangle_{\psi
}$, then the conservation of both symplectic structure and effective energy
are guaranteed automatically.\cite{Ohsawa_Leok:2013} This is because the
function $H_{\text{eff}}(q,p,\mathcal{A},\mathcal{B})$ provides a Hamiltonian
function on a manifold with coordinates $\Lambda$ and symplectic structure
(\ref{eq:omega_Heller_reduced}) and because every Hamiltonian flow conserves
its energy and symplectic structure. Although most examples in the following
Secs.~\ref{sec:TG_methods} and \ref{sec:FG_methods} satisfy $H_{\text{eff}%
}=\langle\hat{H}_{\text{appr}}\rangle_{\psi}$, we do not assume the validity
of this relation. Therefore, the more general Eqs.~(\ref{eq:q_dot}%
)--(\ref{eq:gamma_dot}) obtained from the perspective of the nonlinear
TDSE\ differ from the equations of motion obtained from the Hamiltonian
approach and conservation of neither the symplectic form nor the effective
energy is guaranteed.

The analysis of symplectic structure of Gaussian wavepacket dynamics with a
general effective potential~(\ref{eq:V_eff}) can be done elegantly using the
formalism of symplectic geometry, as was done by Ohsawa and
Leok\cite{Ohsawa_Leok:2013} for the effective potential obtained as the
expected value of some approximate potential ($V_{\text{eff}}:=\langle\hat
{V}_{\text{appr}}\rangle$). Because this analysis relies on nonelementary
concepts of differential geometry, it will be presented
elsewhere.\cite{Vanicek:symplecticity}

\section{\label{sec:TG_methods}Thawed Gaussian wavepacket dynamics}

The reader may ask whether there exist any interesting effective quadratic
Hamiltonians, for which the preceding general analysis is useful. Indeed,
there are many; five of such Hamiltonians are hidden behind the
\emph{variational Gaussian approximation}%
,\cite{Heller:1976,Coalson_Karplus:1990} Heller's original \emph{thawed
Gaussian approximation},\cite{Heller:1975} \emph{single-Hessian thawed
Gaussian approximation},\cite{Begusic_Vanicek:2019} \emph{local cubic
variational TGWD}\ (also known as the \emph{extended semiclassical
dynamics}\cite{Pattanayak_Schieve:1994a} or \emph{symplectic semiclassical
wavepacket dynamics}\cite{Ohsawa:2015,Ohsawa:2015a}), and---of course---the
\emph{global harmonic approximation}, of which the last one leads to a linear
TDSE, while the first four give rise to genuine nonlinear TDSEs. Below, we
also propose a \emph{single-quartic variational TGWD}, which improves the
accuracy of the local cubic variational TGWD without increasing its cost and
without sacrificing its geometric properties.

We now list the expansion coefficients $V_{0}$, $V_{1}$, $V_{2}$ of the
effective potential (\ref{eq:V_eff}) for each of these approximations.
Equations of motion for parameters $q_{t}$, $p_{t}$, $A_{t}$, and $\gamma_{t}$
are, in each case, obtained by substituting specific expressions for $V_{0}$,
$V_{1}$, $V_{2}$ into the general Eqs.~(\ref{eq:q_dot})-(\ref{eq:gamma_dot}).

In this section, we will discuss methods employing a \textquotedblleft
thawed\textquotedblright\ Gaussian---a Gaussian wavepacket with a flexible
width matrix, while in Sec.~\ref{sec:FG_methods}, we shall provide examples of
methods using a \textquotedblleft frozen\textquotedblright\ Gaussian---a
Gaussian wavepacket with a time-independent width.

\subsection{\label{sec:V_TGWD}Variational TGWD}

As shown in
Refs.~\onlinecite{Coalson_Karplus:1990,book_Lubich:2008,Lasser_Lubich:2020}
and here in Appendix~\ref{sec:VGA}, the optimal solution (in the sense of the
Dirac-Frenkel variational
principle\cite{Dirac:1930,book_Frenkel:1934,book_Lubich:2008}) of the TDSE
(\ref{eq:tdse}) with a Gaussian ansatz (\ref{eq:GWP})\ is the
\emph{variational TGWD} or variational Gaussian
approximation,\cite{Heller:1976,Coalson_Karplus:1990,book_Lubich:2008,Lasser_Lubich:2020,Werther_Grossmann:2021}
which corresponds to an effective potential~(\ref{eq:V_eff}) with coefficients%
\begin{equation}
V_{0}=\langle\hat{V}\rangle-\operatorname{Tr}[\langle\hat{V}^{\prime\prime
}\rangle\cdot\Sigma_{t}]/2\text{, }V_{1}=\langle\hat{V}^{\prime}\rangle\text{,
}V_{2}=\langle\hat{V}^{\prime\prime}\rangle. \label{eq:V_VGA}%
\end{equation}
Inserting $V_{j}$ from Eq.~(\ref{eq:V_VGA}) into general
Eqs.~(\ref{eq:exp_value_V_eff})--(\ref{eq:exp_value_Hess_V_eff}) shows that
the variational TGWD preserves expectation values of the potential energy,
gradient, and Hessian:%
\begin{equation}
\langle\hat{V}_{\text{eff}}\rangle=\langle\hat{V}\rangle\text{, }\langle
\hat{V}_{\text{eff}}^{\prime}\rangle=\langle\hat{V}^{\prime}\rangle\text{,
}\langle\hat{V}_{\text{eff}}^{\prime\prime}\rangle=\langle\hat{V}%
^{\prime\prime}\rangle. \label{eq:exp_V_j_VGA}%
\end{equation}
The first equality results from a beautiful cancellation of two terms in the
expression%
\begin{align}
\langle\hat{V}_{\text{eff}}\rangle &  =\langle\hat{V}\rangle-\operatorname{Tr}%
[\langle\hat{V}^{\prime\prime}\rangle\cdot\Sigma_{t}]/2+\operatorname{Tr}%
[\langle\hat{V}^{\prime\prime}\rangle\cdot\Sigma_{t}]/2\nonumber\\
&  =\langle\hat{V}\rangle, \label{eq:exp_V_VGA}%
\end{align}
and implies that the effective energy is exact ($E_{\text{eff}}=E$) for
variational TGWD even though the propagation itself may be far from exact.
Equations of motion that follow from the effective potential~(\ref{eq:V_VGA})
were originally derived (differently) by Coalson and
Karplus\cite{Coalson_Karplus:1990} and are equivalent to those of Theorem 3.2
by Ohsawa and Leok \cite{Ohsawa_Leok:2013} and Theorem 3.11 by Lasser and
Lubich.\cite{Lasser_Lubich:2020} Poirier derived these equations using quantum
trajectories.\cite{Poirier:2017}

The variational TGWD is symplectic.\cite{Faou_Lubich:2006,Ohsawa_Leok:2013}
Because any solution derived from the Dirac-Frenkel variational principle
conserves energy (see Appendix~\ref{sec:cons_E_norm_by_DFVP}%
)\cite{book_Lubich:2008,Lasser_Lubich:2020} and because the exact and
effective energies are equal in the variational TGWD, this approximation
conserves the effective energy, too. Conservation of the exact and effective
energies by the variational TGWD also follows directly from the general
expressions~(\ref{eq:dE_dt_3}) for $dE/dt$ and (\ref{eq:d_exp_Heff_dt}) for
$dE_{\text{eff}}/dt$. See Appendix~\ref{sec:conservation_E_VGA} for this more
\textquotedblleft pedestrian\textquotedblright\ proof of $\dot{E}_{\text{eff}%
}=0$.

The variational TGWD has been extended from real-time to imaginary-time
quantum dynamics in order to describe equilibrium properties of van der Waals
clusters\cite{Frantsuzov_Mandelshtam:2004} and time-correlation functions of
liquid \textit{para}-hydrogen.\cite{Georgescu_Deckman:2011}

\subsection{Local harmonic TGWD}

In his original thawed Gaussian approximation,\cite{Heller:1975,Heller:1981a}
Heller did not invoke the variational principle and avoided the expensive
evaluation of expectation values needed in Eq. (\ref{eq:V_VGA}) by making the
\emph{local harmonic approximation}, in which the effective potential in
Eq.~(\ref{eq:V_eff}) depends on $\psi$ only via $q_{t}$ and its coefficients%
\begin{equation}
V_{0}=V\left(  q_{t}\right)  \text{, }V_{1}=V^{\prime}\left(  q_{t}\right)
\text{, }V_{2}=V^{\prime\prime}\left(  q_{t}\right)  \label{eq:V_LHA_1}%
\end{equation}
are the coefficients of the truncated, second-order Taylor expansion of
$V(\hat{q})$ about $q_{t}$. Heller's local harmonic TGWD (\ref{eq:V_LHA_1})
can be also obtained from the variational TGWD (\ref{eq:V_VGA}) if the local
harmonic approximation is used to evaluate the expectation values $\langle
\hat{V}^{(j)}\rangle$. In Sec.~\ref{sec:VGA_w_V_appr}, we prove this statement
for an arbitrary \textquotedblleft local quadratic\textquotedblright%
\ approximation for $V$.

If we substitute expressions for $V_{0}$ and $V_{1}$ from
Eq.~(\ref{eq:V_LHA_1}) into the general Eq.~(\ref{eq:exp_value_V_eff}), we
find that%
\begin{equation}
\langle\hat{V}_{\text{eff}}\rangle=V\left(  q_{t}\right)  +\operatorname{Tr}%
\left[  V^{\prime\prime}\left(  q_{t}\right)  \cdot\Sigma_{t}\right]
/2\neq\langle\hat{V}\rangle
\end{equation}
and, therefore, $E_{\text{eff}}\neq E$. The local harmonic TGWD conserves
neither the exact nor the effective energy. Whereas the nonconservation of the
exact energy was proven in general in the discussion following
Eq.~(\ref{eq:dE_dt_3}) in Sec.~\ref{subsec:time_dependence_of_E}, the
nonconservation of the effective energy follows from
Eq.~(\ref{eq:d_exp_Heff_dt}) because%
\begin{align}
\dot{E}_{\text{eff}}  &  =V^{\prime}(q_{t})^{T}\cdot\dot{q}_{t}-V^{\prime
}(q_{t})^{T}\cdot\dot{q}_{t}+\operatorname{Tr}\left(  B_{t}\cdot\Sigma
_{t}\right)  /2\nonumber\\
&  =\operatorname{Tr}\left(  B_{t}\cdot\Sigma_{t}\right)  /2,
\end{align}
where $B_{t}$ is a matrix obtained from the contraction of vector $\dot{q}%
_{t}$ with the symmetric rank-$3$ tensor $V^{\prime\prime\prime}(q_{t})$.
Using Einstein's convention for a sum over repeated indices,
\begin{align}
\left(  B_{t}\right)  _{jk}  &  :=dV^{\prime\prime}(q_{t})_{jk}/dt=\left(
\dot{q}_{t}\right)  _{i}V^{\prime\prime\prime}(q_{t})_{ijk}\nonumber\\
&  =(p_{t})_{l}(m^{-1})_{li}V^{\prime\prime\prime}(q_{t})_{ijk}.
\label{eq:B_t}%
\end{align}

Lauvergnat et al.\cite{Lauvergnat_Desouter-Lecomte:2006} derived equations of
motion of the local harmonic TGWD in generalized coordinates.

\subsection{\label{sec:SHA}Single-Hessian TGWD}

The most expensive part of a higher-dimensional calculation using the local
harmonic TGWD is, of course, the evaluation of the Hessian matrix
$V^{\prime\prime}\left(  q_{t}\right)  $. In the \emph{single-Hessian
approximation}%
,\cite{Begusic_Vanicek:2019,Begusic_Vanicek:2021,Begusic_Vanicek:2021a} the
Hessian is computed only once, at a reference geometry $q_{\text{ref}}$, but
the energies and gradients are still computed at each point $q_{t}$ along the
trajectory. Within the \emph{single-Hessian TGWD}, the effective
potential~(\ref{eq:V_eff}) again depends on $\psi$ only via $q_{t}$:%
\begin{equation}
V_{0}=V\left(  q_{t}\right)  \text{, }V_{1}=V^{\prime}\left(  q_{t}\right)
\text{, }V_{2}=V^{\prime\prime}(q_{\text{ref}})\text{.} \label{eq:V_SHA}%
\end{equation}
Although the single-Hessian TGWD does not conserve energy, it is symplectic
and Eq.~(\ref{eq:d_exp_Heff_dt}) implies the conservation of the effective
energy:\cite{Begusic_Vanicek:2019}
\begin{align}
\dot{E}_{\text{eff}}  &  =V^{\prime}(q_{t})^{T}\cdot\dot{q}_{t}-V^{\prime
}(q_{t})^{T}\cdot\dot{q}_{t}+\frac{1}{2}\operatorname{Tr}\left[
\frac{dV^{\prime\prime}(q_{\text{ref}})}{dt}\cdot\Sigma_{t}\right] \nonumber\\
&  =0
\end{align}
because $dV^{\prime\prime}(q_{\text{ref}})/dt=0$ as $q_{\text{ref}}$ is
constant. Because of its efficiency and improved geometric properties, the
single-Hessian TGWD was implemented in the electronic structure package
Turbomole.\cite{Begusic_Vanicek:2022}

\subsection{Global harmonic TGWD}

Among all thawed Gaussian approximations, the least expensive but crudest one
is the \emph{global harmonic TGWD}, in which the effective potential is the
second-order Taylor expansion of $V$ about a fixed reference geometry
$q_{\text{ref}}$:%
\begin{align}
V_{\text{eff}}(q)  &  =V(q_{\text{ref}})+V^{\prime}(q_{\text{ref}}){}^{T}%
\cdot(q-q_{\text{ref}})\nonumber\\
&  ~~~+(q-q_{\text{ref}})^{T}\cdot V^{\prime\prime}(q_{\text{ref}}%
)\cdot(q-q_{\text{ref}})/2. \label{eq:V_HA}%
\end{align}
This equation is not in the standard form (\ref{eq:V_eff}), which requires an
expansion about the current center $q_{t}$ of the wavepacket; the coefficients
$V_{j}$ of the standard form (\ref{eq:V_eff}) are obtained by evaluating
derivatives $V_{\text{eff}}^{(j)}$ at $q_{t}$:%
\begin{align}
V_{0}  &  =V_{\text{eff}}(q_{t}),\nonumber\\
V_{1}  &  =V_{\text{eff}}^{\prime}(q_{t})=V^{\prime}(q_{\text{ref}}%
)+V^{\prime\prime}(q_{\text{ref}})\cdot(q_{t}-q_{\text{ref}}%
),\label{eq:V_HA_std}\\
V_{2}  &  =V_{\text{eff}}^{\prime\prime}(q_{t})=V^{\prime\prime}%
(q_{\text{ref}}).\nonumber
\end{align}
Because the coefficients $V_{0}$ and $V_{1}$ depend on $\psi$ via $q_{t}$, one
might think that $V_{\text{eff}}$ is a nonlinear operator. Yet, in contrast to
the previously mentioned approximations, in the global harmonic TGWD
$V_{\text{eff}}$ is a \emph{linear} operator; this follows clearly from
Eq.~(\ref{eq:V_HA}), where $V_{\text{eff}}$ depends on $\psi$ via neither
$q_{t}$ nor any other parameter of the Gaussian.

Although the global harmonic TGWD does not conserve energy, it obviously
conserves both the symplectic structure and effective energy $E_{\text{eff}}$
because $\hat{H}_{\text{eff}}=\hat{T}+\hat{V}_{\text{eff}}$ is a linear
time-independent Hamiltonian operator. An alternative proof follows from
Eq.~(\ref{eq:d_exp_Heff_dt}):
\begin{align}
\dot{E}_{\text{eff}}  &  =[V^{\prime}(q_{\text{ref}})+V^{\prime\prime
}(q_{\text{ref}})\cdot(q_{t}-q_{\text{ref}})]^{T}\cdot\dot{q}_{t}\nonumber\\
&  ~~~-[V^{\prime}(q_{\text{ref}})+V^{\prime\prime}(q_{\text{ref}})\cdot
(q_{t}-q_{\text{ref}})]^{T}\cdot\dot{q}_{t}=0.
\end{align}

\subsection{\label{sec:VGA_w_V_appr}Variational Gaussian approximation applied
to any local quadratic approximation for $V$}

Effective potentials used in the local harmonic, single-Hessian, and global
harmonic TGWD are all quadratic functions of nuclear coordinates and, as a
result, can be obtained either directly (as suggested above) or by an
alternative procedure, consisting of two steps: First, approximate the exact
potential $V$ by a state-dependent approximation $V_{\text{appr}}$. Then,
apply the variational TGWD to $V_{\text{appr}}$ instead of $V$. To see this,
note that if%
\begin{equation}
V_{\text{appr}}(x)=v_{0}+v_{1}^{T}\cdot x+x^{T}\cdot v_{2}\cdot x/2
\label{eq:V_approx}%
\end{equation}
is a quadratic polynomial of $x$, inserting $V_{\text{appr}}$ instead of $V$
into the variational expressions~(\ref{eq:V_VGA}) for $V_{j}$ yields
\begin{align}
V_{0}  &  =\langle\hat{V}_{\text{appr}}\rangle-\operatorname{Tr}[\langle
\hat{V}_{\text{appr}}^{\prime\prime}\rangle\cdot\Sigma_{t}]/2\nonumber\\
&  =v_{0}+\frac{1}{2}\operatorname{Tr}(v_{2}\cdot\Sigma_{t})-\frac{1}%
{2}\operatorname{Tr}(v_{2}\cdot\Sigma_{t})=v_{0},\\
V_{1}  &  =\langle\hat{V}_{\text{appr}}^{\prime}\rangle=v_{1},\\
V_{2}  &  =\langle\hat{V}_{\text{appr}}^{\prime\prime}\rangle=v_{2},
\end{align}
where the expectation values were evaluated by applying
Eqs.~(\ref{eq:exp_value_V_eff})--(\ref{eq:exp_value_Hess_V_eff}) to
$V_{\text{appr}}$ instead of $V_{\text{eff}}$. The effective potential is
equal to the original approximate potential:%
\begin{equation}
V_{\text{eff}}(x)=V_{\text{appr}}(x)\text{.} \label{eq:V_eff_V_appr}%
\end{equation}
Expectation values of the effective potential, its gradient and Hessian are,
therefore, equal to the corresponding properties of the approximate potential:%
\begin{align}
\langle\hat{V}_{\text{eff}}\rangle &  =\langle\hat{V}_{\text{appr}}%
\rangle=v_{0}+\operatorname{Tr}(v_{2}\cdot\Sigma_{t})/2\text{,\ }\\
\langle\hat{V}_{\text{eff}}^{\prime}\rangle &  =\langle\hat{V}_{\text{appr}%
}^{\prime}\rangle=v_{1}\text{,\ }\\
\langle\hat{V}_{\text{eff}}^{\prime\prime}\rangle &  =\langle\hat
{V}_{\text{appr}}^{\prime\prime}\rangle=v_{2}.
\end{align}
The variational principle is not needed if the effective potential is
\emph{defined} by Eq.~(\ref{eq:V_eff_V_appr}). Equality of the effective and
exact energies requires that $v_{0}=\langle\hat{V}\rangle-\operatorname{Tr}%
(v_{2}\cdot\Sigma_{t})/2$. This equality holds in the variational TGWD, where
$v_{2}=\langle\hat{V}^{\prime\prime}\rangle$. If $v_{0}$ only depends on
$q_{t}$, Eq.~(\ref{eq:d_exp_Heff_dt}) implies that%
\[
\dot{E}_{\text{eff}}=[v_{0}^{\prime}(q_{t})-v_{1}]^{T}\cdot\dot{q}%
_{t}+\operatorname{Tr}(\dot{v}_{2}\cdot\Sigma_{t})/2.
\]
This shows that if the variational principle is applied to an approximate
instead of the exact Hamiltonian, neither the exact nor effective energy is
conserved in general. The case of local harmonic TGWD demonstrates that the
symplectic structure is generally not conserved, either.

\subsection{\label{sec:LCV_TGWD}Local cubic variational TGWD}

Compared with Heller's original TGWD, the variational TGWD is hard to evaluate
in practice because expectation values $\langle\hat{V}^{(j)}\rangle$ typically
cannot be obtained analytically. However, as we have just shown, to go beyond
Heller's method, the variational TGWD must be combined with a more accurate
than a local harmonic approximation for $V$. Then the effective quadratic
potential $V_{\text{eff}}$ will, obviously, differ from the approximate
potential $V_{\text{appr}}$. An obvious and the simplest possible choice for
$V_{\text{appr}}$ that still permits evaluating the expectation values
analytically is the \emph{local cubic approximation}%
\begin{align}
V_{\text{appr}}(q)  &  :=V(q_{t})+V^{\prime}(q_{t})^{T}\cdot x+x^{T}\cdot
V^{\prime\prime}(q_{t})\cdot x/2\nonumber\\
&  ~~~+V^{\prime\prime\prime}\left(  q_{t}\right)  _{ijk}x_{i}x_{j}x_{k}/3!,
\label{eq:V_LCA}%
\end{align}
which was also used for evaluating the matrix elements of the potential in the
variational multi-configurational Gaussian method.\cite{Bonfanti_Pollak:2018}
Expectation values of $V_{\text{appr}}$ are%
\begin{align}
\langle\hat{V}_{\text{appr}}^{\prime\prime}\rangle &  =V^{\prime\prime}%
(q_{t}),\\
\langle\hat{V}_{\text{appr}}^{\prime}\rangle_{i}  &  =V^{\prime}(q_{t}%
)_{i}+V^{\prime\prime\prime}(q_{t})_{ijk}\Sigma_{t,jk}/2,\\
\langle\hat{V}_{\text{appr}}\rangle &  =V(q_{t})+\operatorname{Tr}%
[V^{\prime\prime}(q_{t})\cdot\Sigma_{t}]/2.
\end{align}
Substitution of these into the variational expressions~(\ref{eq:V_VGA}) for
$V_{j}$ yields the effective potential coefficients%
\begin{align}
V_{0}  &  =V(q_{t}),\nonumber\\
V_{1,i}  &  =V^{\prime}(q_{t})_{i}+V^{\prime\prime\prime}(q_{t})_{ijk}%
\Sigma_{t,jk}/2,\text{ }\label{eq:V_VGA-LCA}\\
V_{2}  &  =V^{\prime\prime}(q_{t})\nonumber
\end{align}
of the \emph{local cubic variational TGWD}. Equations of motion obtained from
the effective potential~(\ref{eq:V_VGA-LCA}) are equivalent to those of
\textquotedblleft symplectic semiclassical wavepacket
dynamics\textquotedblright\ [Eq.~(36) in Ref.~\onlinecite{Ohsawa_Leok:2013},
Eq.~(19) in Ref.~\onlinecite{Ohsawa:2015}, and Proposition 4.4 in
Ref.~\onlinecite{Ohsawa:2015a}] and, indirectly, with those of
\textquotedblleft extended semiclassical dynamics\textquotedblright\ of
Pattanayak and Schieve.\cite{Pattanayak_Schieve:1994a} The method is
symplectic\cite{Ohsawa_Leok:2013} and only the equation for $\dot{p}_{t}$
differs from the local harmonic TGWD due to a nonclassical term in the force,
given by $-V_{1}$.

Equation~(\ref{eq:d_exp_Heff_dt}) implies that the local cubic variational
TGWD conserves the effective energy because%
\begin{align}
\dot{E}_{\text{eff}}  &  =\dot{V}_{0}-V_{1}^{T}\cdot\dot{q}_{t}%
+\operatorname{Tr}(\dot{V}_{2}\cdot\Sigma_{t})/2\nonumber\\
&  =\dot{V}(q_{t})-V^{\prime}(q_{t})^{T}\cdot\dot{q}_{t}-V^{\prime\prime
\prime}(q_{t})_{ijk}\Sigma_{t,jk}\dot{q}_{t},_{i}/2\nonumber\\
&  ~~~+\operatorname{Tr}\{[dV^{\prime\prime}(q_{t})/dt]\cdot\Sigma_{t}\}/2=0,
\end{align}
where, in the last step, the second and fourth terms cancel the first and the
third terms [see Eq.~(\ref{eq:B_t})].

\subsection{\label{sec:SQV-TGWD}Single-quartic variational TGWD (see
Fig.~\ref{fig:GWP_and_V_eff})}

To increase the accuracy over the local cubic approximation, the
\textquotedblleft obvious\textquotedblright logical step is to include the
local fourth derivative of $V$. However, evaluating this \emph{local quartic
approximation} is expensive and, like the local harmonic TGWD, the \emph{local
quartic variational TGWD} conserves neither the effective energy nor
symplectic structure. Instead, in analogy to the single-Hessian TGWD, a much
more effective approach is applying the variational TGWD to the
\emph{single-quartic approximation}%
\begin{align}
V_{\text{appr}}(q)  &  :=V(q_{t})+V^{\prime}(q_{t})^{T}\cdot x+x^{T}\cdot
V^{\prime\prime}(q_{t})\cdot x/2\nonumber\\
&  ~~~+V^{\prime\prime\prime}\left(  q_{t}\right)  _{ijk}x_{i}x_{j}%
x_{k}/3!\nonumber\\
&  ~~~+V^{(4)}(q_{\text{ref}})_{ijkl}x_{i}x_{j}x_{k}x_{l}/4!, \label{eq:V_SQA}%
\end{align}
which augments the local cubic approximation (\ref{eq:V_LCA}) with the
evaluation of a single fourth derivative at a reference geometry
$q_{\text{ref}}$. Coefficients of the effective potential of this
\emph{single-quartic variational TGWD }are obtained by applying the
variational formulas (\ref{eq:V_VGA}) to the expectation values of the
potential, gradient, and Hessian of $V_{\text{appr}}$:%
\begin{align}
V_{2,}{}_{ij}  &  =\langle\hat{V}_{\text{appr}}^{\prime\prime}\rangle
_{ij}=V^{\prime\prime}(q_{t})_{ij}+V^{(4)}(q_{\text{ref}})_{ijkl}\Sigma
_{t,kl}/2,\label{eq:V2_SQV-TGWD}\\
V_{1,i}  &  =\langle\hat{V}_{\text{appr}}^{\prime}\rangle_{i}=V^{\prime}%
(q_{t})_{i}+V^{\prime\prime\prime}(q_{t})_{ijk}\Sigma_{t,jk}%
/2,\label{eq:V1_SQV-TGWD}\\
\langle\hat{V}_{\text{appr}}\rangle &  =V(q_{t})+\operatorname{Tr}%
[V^{\prime\prime}(q_{t})\cdot\Sigma_{t}]/2\nonumber\\
&  ~~~+V^{(4)}(q_{\text{ref}})_{ijkl}\Sigma_{t,ij}\Sigma_{t,kl}%
/8,\label{eq:exp_value_V_SQA}\\
V_{0}  &  =\langle\hat{V}_{\text{appr}}\rangle-\operatorname{Tr}[\langle
\hat{V}_{\text{appr}}^{\prime\prime}\rangle\cdot\Sigma_{t}]/2\nonumber\\
&  =V(q_{t})-V^{(4)}(q_{\text{ref}})_{ijkl}\Sigma_{t,ij}\Sigma_{t,kl}/8.
\label{eq:V0_SQV-TGWD}%
\end{align}
Here, in deriving expressions for $\langle\hat{V}_{\text{appr}}^{\prime
}\rangle$ and $\langle\hat{V}_{\text{appr}}\rangle$, we used the identities
$\langle x_{i}x_{j}x_{k}\rangle=0$ and%
\begin{equation}
\langle x_{i}x_{j}x_{k}x_{l}\rangle=\Sigma_{t,ij}\Sigma_{t,kl}+\Sigma
_{t,ik}\Sigma_{t,jl}+\Sigma_{t,il}\Sigma_{t,jk};
\end{equation}
since $V^{(4)}(q_{\text{ref}})_{ijkl}$ is a totally symmetric tensor,%
\[
V^{(4)}(q_{\text{ref}})_{ijkl}\langle x_{i}x_{j}x_{k}x_{l}\rangle
=3V^{(4)}(q_{\text{ref}})_{ijkl}\Sigma_{t,ij}\Sigma_{t,kl}.
\]
In the single-quartic variational TGWD, equations for $\dot{q}_{t}$ and
$\dot{p}_{t}$ remain the same as in the local cubic variational TGWD, but
equations for $\dot{A}_{t}$ and $\dot{\gamma}_{t}$ change due to the changes
in $V_{0}$ and $V_{2}$. Because only a single fourth derivative is needed, the
computational cost is only slightly higher than the cost of the local cubic
variational TGWD: assuming that the fourth derivative must be evaluated by
finite difference, the increase in cost is negligible in typical simulations,
where the number of time steps is much larger than the number of degrees of freedom.

Remarkably, Eq.~(\ref{eq:d_exp_Heff_dt}) implies that the single-quartic
variational TGWD conserves the effective energy because%
\begin{align}
\dot{E}_{\text{eff}}  &  =\dot{V}_{0}-V_{1}^{T}\cdot\dot{q}_{t}%
+\operatorname{Tr}(\dot{V}_{2}\cdot\Sigma_{t})/2\nonumber\\
&  =\dot{V}(q_{t})-V^{(4)}(q_{\text{ref}})_{ijkl}\dot{\Sigma}_{t,ij}%
\Sigma_{t,kl}/4\nonumber\\
&  ~~~-V^{\prime}(q_{t})^{T}\cdot\dot{q}_{t}-V^{\prime\prime\prime}%
(q_{t})_{ijk}\Sigma_{t,jk}\dot{q}_{t,i}/2\nonumber\\
&  ~~~+\frac{1}{2}\operatorname{Tr}\left[  \frac{dV^{\prime\prime}(q_{t})}%
{dt}\cdot\Sigma_{t}\right]  +\frac{1}{4}V^{(4)}(q_{\text{ref}})_{ijkl}%
\dot{\Sigma}_{t,ij}\Sigma_{t,kl}\nonumber\\
&  =0,
\end{align}
where, in the last step, the first term cancels the third one, the second term
cancels the sixth, and the fourth term cancels the fifth [see
Eq.~(\ref{eq:B_t})]. Because the single-quartic variational TGWD does not
increase the cost over the local cubic variational TGWD and because it, in
contrast to the local quartic TGWD, conserves the symplectic structure and
effective energy, this method appears to be the logical and promising choice
in calculations whose goal is to improve geometric properties and accuracy
beyond Heller's original local harmonic TGWD.

\section{\label{sec:FG_methods}Frozen Gaussian wavepacket dynamics}

In more restrictive approximations, the Gaussian wavepacket has a constant
width matrix $A_{t}=A_{0}$. Clearly, all such \textquotedblleft frozen
Gaussian\textquotedblright\ approximations\cite{Heller:1981} require that
$\dot{A}_{t}=0$ and Eq.~(\ref{eq:A_dot}) implies that the coefficient $V_{2}$
in the effective quadratic potential (\ref{eq:V_eff}) must satisfy%
\begin{equation}
V_{2}=-A_{t}\cdot m^{-1}\cdot A_{t}=-A_{0}\cdot m^{-1}\cdot A_{0}.
\label{eq:V2_all_FGA}%
\end{equation}
One can choose freely only the coefficients $V_{0}$ and $V_{1}$ but not
$V_{2}$. The frozen Gaussian approximation is typically used in
multi-trajectory methods, which can describe wavepacket spreading and
distortion without requiring a flexible width; the description of the
nonlinear spreading of a wavepacket was, indeed, Heller's motivation in
proposing the frozen Gaussian approximation.\cite{Heller:1981} Other methods
employing frozen Gaussians are the semiclassical Herman--Kluk
propagator\cite{Herman_Kluk:1984,Miller:2001,Lasser_Lubich:2020,Werther_Grossmann:2021}
and its
extensions,\cite{Kaledin_Miller:2003,Ceotto_Conte:2017,Church_Ananth:2017}
which associate to each trajectory a weight factor depending on the stability
matrix, imaginary time frozen Gaussian dynamics,\cite{Cartarius_Pollak:2011}
which treats the quantum Boltzmann operator instead of the time evolution
operator, as well as Gaussian basis
methods,\cite{Martinez_Levine:1997,Child_Shalashilin:2003,Burghardt_Giri:2008,Sulc_Vanicek:2013,Gu_Garashchuk:2016}
which allow coupling between the trajectories.

Here, we focus only on \emph{single-trajectory} methods employing a Gaussian
with a fixed width. These methods form a family of \emph{frozen Gaussian
wavepacket dynamics} (FGWD), which has an interesting relation to the
single-Hessian approximation (\ref{eq:V_SHA}) of Sec.~\ref{sec:SHA}. The FGWD
is, on one hand, a special case of the single-Hessian TGWD because one can
think of $V_{2}$ from Eq.~(\ref{eq:V2_all_FGA}) as a reference Hessian of a
harmonic potential whose ground state is the initial state and, on the other
hand, a generalization because one is still free to choose $V_{0}$ and $V_{1}$.

Let
\begin{align}
\mathcal{A}  &  :=\operatorname{Re}A_{0}=\operatorname{Re}A_{t}\text{
\ \ and}\label{eq:Re_A}\\
\mathcal{B}  &  :=\operatorname{Im}A_{0}=\operatorname{Im}A_{t}
\label{eq:Im_A}%
\end{align}
denote the real and imaginary parts of the constant width matrix. It follows
from Eq.~(\ref{eq:V2_all_FGA}) that
\begin{align}
\operatorname{Re}V_{2}  &  =-\mathcal{A}\cdot m^{-1}\cdot\mathcal{A}%
+\mathcal{B}\cdot m^{-1}\cdot\mathcal{B},\\
\operatorname{Im}V_{2}  &  =-\mathcal{A}\cdot m^{-1}\cdot\mathcal{B}%
-\mathcal{B}\cdot m^{-1}\cdot\mathcal{A}. \label{eq:Im_V2}%
\end{align}
Recall that in order that $\hat{V}_{\text{eff}}(\psi)$ be a Hermitian operator
with real expectation values, coefficient $V_{2}(\psi)$ must be a real matrix.
In Appendix$~$\ref{app:zero_ReA}, it is shown that setting $\operatorname{Im}%
V_{2}=0$ in Eq.~(\ref{eq:Im_V2}) implies that the initial width matrix $A_{0}$
of the frozen Gaussian is purely imaginary:%
\begin{align}
{\mathcal{A}}  &  =0{}\text{ \ \ and}\label{eq:zero_ReA}\\
V_{2}  &  =\operatorname{Re}V_{2}=\mathcal{B}\cdot m^{-1}\cdot\mathcal{B}.
\label{eq:V2_FGWD}%
\end{align}

Because $\mathcal{A}=0$ in the FGWD, $\operatorname{Cov}_{R}(\hat{q},\hat
{p})=0$ and the equations of motion~(\ref{eq:q_dot})-(\ref{eq:gamma_dot})
become%
\begin{align}
\dot{q}_{t}  &  =m^{-1}\cdot p_{t},\\
\dot{p}_{t}  &  =-V_{1},\\
A_{t}  &  =A_{0}=\operatorname{const}=i\mathcal{B},\label{eq:A0_purely_imag}\\
\dot{\gamma}_{t}  &  =T(p_{t})-V_{0}-(\hbar/2)\operatorname*{Tr}\left(
m^{-1}\cdot\mathcal{B}\right)  .
\end{align}
Equations~(\ref{eq:dE_dt_3}) and (\ref{eq:d_exp_Heff_dt}) for the time
dependence of exact and effective energies reduce to%
\begin{align}
\dot{E}  &  =p_{t}^{T}\cdot m^{-1}\cdot(\langle\hat{V}^{\prime}\rangle
-V_{1})\text{ \ \ and}\label{eq:dE_dt_FGWD}\\
\dot{E}_{\text{eff}}  &  =\dot{V}_{0}-V_{1}^{T}\cdot\dot{q}_{t}=\dot{V}%
_{0}+\dot{p}_{t}^{T}\cdot m^{-1}\cdot p_{t}\nonumber\\
&  =\dot{V}_{0}+\dot{T}(p_{t}). \label{eq:dEeff_dt_FGWD}%
\end{align}
Now, let us examine examples of effective potentials giving rise to FGWD;
$V_{2}$ is always given by Eq.~(\ref{eq:V2_FGWD}).

\subsection{\label{sec:V_FGWD}Variational FGWD}

Application of the Dirac-Frenkel variational principle to the frozen Gaussian
ansatz is similar to the derivation of the variational TGWD in
Appendix~\ref{sec:VGA}\ except that we lose parameter $A_{t}$ and, therefore,
Eqs.~(\ref{eq:DFVP_A}) and (\ref{eq:DFVP_A_exp}). The remaining equations are
either $0=\left\langle g\right\rangle =\left\langle xg\right\rangle $ or the
equivalent $0=\left\langle g\right\rangle =\left\langle \nabla g\right\rangle
$, which give%
\begin{align}
V_{0}  &  =\langle\hat{V}\rangle-\operatorname{Tr}\left[  V_{2}\cdot
\operatorname{Cov}(\hat{q})\right]  /2\nonumber\\
&  =\langle\hat{V}\rangle+(i\hbar/4)\operatorname{Tr}\left(  A_{0}\cdot
m^{-1}\cdot A_{0}\cdot A_{0}^{-1}\right) \nonumber\\
&  =\langle\hat{V}\rangle-(\hbar/4)\operatorname{Tr}(m^{-1}\cdot
\mathcal{B}),\nonumber\\
V_{1}  &  =\langle\hat{V}^{\prime}\rangle. \label{eq:VFGA}%
\end{align}
Equation~(\ref{eq:gamma_dot}) for $\dot{\gamma}_{t}$ becomes%
\begin{align}
\dot{\gamma}_{t}  &  =T(p_{t})-\langle\hat{V}\rangle-(\hbar
/4)\operatorname{Tr}\left(  m^{-1}\cdot\mathcal{B}\right) \nonumber\\
&  =T(p_{t})-\langle\hat{V}\rangle-\operatorname{Tr}\left[  m^{-1}%
\cdot\operatorname{Cov}(\hat{p})\right]  /2\nonumber\\
&  =2T(p_{t})-\langle\hat{V}\rangle-\langle\hat{T}\rangle=2T(p_{t}%
)-\langle\hat{H}\rangle.
\end{align}
Since $V_{1}=\langle\hat{V}^{\prime}\rangle$ and $E_{\text{eff}}=E$ in the
variational FGWD, Eq.~(\ref{eq:dE_dt_FGWD}) implies that the exact and
effective energies are conserved: $\dot{E}=\dot{E}_{\text{eff}}=0$.

\subsection{Variational FGWD with classical trajectories}

In Heller's original frozen Gaussian approximation,\cite{Heller:1981} the
Gaussians move along classical trajectories. Such an approximation can be
obtained by applying the Dirac-Frenkel variational principle to the Gaussian
ansatz, in which $q_{t}$ and $p_{t}$ are required to follow classical
Hamilton's equations of motion and $\gamma_{t}$ is the only variationally
optimized parameter. (Strictly speaking, this is a generalization of the
Dirac-Frenkel variational principle because the approximation manifold is
time-dependent.) The derivation is analogous to that of the variational TGWD
except that now we only have Eq.~(\ref{eq:DFVP_gamma})\ related to the
parameter $\gamma_{t}$: $0=\left\langle g\right\rangle $. The resulting method
corresponds to an effective potential with coefficients%
\begin{equation}
V_{0}=\langle\hat{V}\rangle-(\hbar/4)\operatorname{Tr}(m^{-1}\cdot
\mathcal{B})\text{ \ and \ }V_{1}=V^{\prime}(q_{t}). \label{eq:FGA}%
\end{equation}
As in the previous case, the exact and effective energies are equal,
$E=E_{\text{eff}}$; however, neither energy is conserved and their common time
dependence is%
\begin{equation}
\dot{E}=\dot{E}_{\text{eff}}=p_{t}^{T}\cdot m^{-1}\cdot\lbrack\langle\hat
{V}^{\prime}\rangle-V^{\prime}(q_{t})].
\end{equation}

\subsection{Variational frozen Gaussian approximation applied to any local
quadratic approximation for $V$}

In analogy to the TGWD, one may apply the variational frozen Gaussian
approximation (\ref{eq:VFGA}) to an approximate, state-dependent potential
$V_{\text{appr}}$ from Eq.~(\ref{eq:V_approx}) instead of $V$. The only
difference is that for the frozen Gaussian approximation, the coefficient
$v_{2}$ of $V_{\text{appr}}$ is fixed and always equal to $V_{2}$ from
Eq.~(\ref{eq:V2_FGWD}). Derivation from Sec.~\ref{sec:VGA_w_V_appr} combined
with the variational approximation (\ref{eq:VFGA}) yields an effective
potential (\ref{eq:V_eff})\ with%
\begin{equation}
V_{0}=v_{0}\text{ \ \ and \ \ }V_{1}=v_{1}. \label{eq:V0_V1_FG_quad}%
\end{equation}
Neither the exact nor the effective energy is conserved in general; their time
dependences are%
\begin{equation}
\dot{E}=p_{t}^{T}\cdot m^{-1}\cdot(\langle\hat{V}^{\prime}\rangle-v_{1})
\end{equation}
and, if the coefficients $v_{0}$ and $v_{1}$ only depend on $q_{t}$,%
\begin{equation}
\dot{E}_{\text{eff}}=\dot{v}_{0}-v_{1}^{T}\cdot\dot{q}_{t}=\left[
v_{0}^{\prime}(q_{t})-v_{1}(q_{t})\right]  ^{T}\cdot\dot{q}_{t}.
\label{eq:E_eff_dot_VFGA_local_quadratic}%
\end{equation}

\subsection{Local harmonic FGWD = single-Hessian FGWD}

Since the coefficient $v_{2}$ of $V_{\text{appr}}$ is always equal to $V_{2}$
from Eq.~(\ref{eq:V2_FGWD}), the local harmonic and single-Hessian
approximations applied to the frozen Gaussian are equivalent; effective
potential coefficients from Eq.~(\ref{eq:V0_V1_FG_quad}) are%
\begin{equation}
V_{0}=v_{0}=V(q_{t})\text{ \ \ and \ \ }V_{1}=v_{1}=V^{\prime}(q_{t}).
\label{eq:FGA_LHA}%
\end{equation}
The same method is obtained regardless of whether the FGWD assumes classical
or variational trajectories. Whereas the exact energy generally depends on
time,%
\begin{equation}
\dot{E}=p_{t}^{T}\cdot m^{-1}\cdot\lbrack\langle\hat{V}^{\prime}%
\rangle-V^{\prime}(q_{t})],
\end{equation}
Eq.~(\ref{eq:E_eff_dot_VFGA_local_quadratic}) implies that the effective
energy is conserved:%
\begin{equation}
\dot{E}_{\text{eff}}=[V^{\prime}(q_{t})-V^{\prime}(q_{t})]^{T}\cdot\dot{q}%
_{t}=0.
\end{equation}

\subsection{Global harmonic FGWD}

Likewise, combining the frozen Gaussian ansatz with the global harmonic
approximation yields an approximation equivalent to $V_{\text{eff}}$ with
coefficients%
\begin{equation}
V_{0}=v_{0}=V_{\text{appr}}(q_{t})\text{ \ \ and \ \ }V_{1}=v_{1}%
=V_{\text{appr}}^{\prime}(q_{t}). \label{eq:FGA_HA}%
\end{equation}
This result follows because the Hessian of the global harmonic approximation
is again constrained by Eq.~(\ref{eq:V2_FGWD}) and $V_{\text{appr}}%
^{\prime\prime}(q_{t})=v_{2}=V_{2}$. Whereas the exact energy in general
depends on time,%
\begin{equation}
\dot{E}=p_{t}^{T}\cdot m^{-1}\cdot\lbrack\langle\hat{V}^{\prime}%
\rangle-V_{\text{appr}}^{\prime}(q_{t})],
\end{equation}
Eq.~(\ref{eq:E_eff_dot_VFGA_local_quadratic}) implies conservation of the
effective energy:%
\begin{equation}
\dot{E}_{\text{eff}}=[V_{\text{appr}}^{\prime}(q_{t})-V_{\text{appr}}^{\prime
}(q_{t})]^{T}\cdot\dot{q}_{t}=0.
\end{equation}

\section{\label{sec:integrators}Geometric integrators}

As mentioned in Sec.~\ref{sec:geom_properties}, exact solutions of the
nonlinear TDSE (\ref{eq:tdse_q-rep}) have certain geometric properties,\ such
as norm conservation and time reversibility. For some effective potentials,
the exact solutions also conserve the symplectic structure, energy, or
effective energy. Numerical solution of the nonlinear equation, however,
requires further approximations, including time discretization.
\textquotedblleft Geometric integrators\textquotedblright\ are numerical
algorithms\cite{book_Leimkuhler_Reich:2004,book_Hairer_Wanner:2006} that
preserve some or all geometric properties of the exact solution, regardless of
the discretization time step.

Let us describe geometric integrators for the nonlinear Schr\"{o}dinger
equation (\ref{eq:tdse_q-rep}) with a general quadratic effective potential
(\ref{eq:V_eff}). Because the effective Hamiltonian (\ref{eq:H_eff}) is
separable into a kinetic energy term depending only on $\hat{p}$ and potential
term depending only on $\hat{q}$, we can, under a rather weak additional
assumption on $V_{\text{eff}}$ (see Sec.~\ref{subsec:V_prop}) employ the
explicit splitting method. In this method, equations of motion are solved
analytically for both the kinetic and potential propagation steps, in which
$\hat{H}_{\text{eff}}:=T(\hat{p})$ and $\hat{H}_{\text{eff}}:=V_{\text{eff}%
}(\hat{q};\psi)$, respectively.

By composing exactly solved kinetic and potential propagations with the same
time step $\Delta t$, one obtains---depending on the ordering of
composition---either the \textquotedblleft VT\textquotedblright\ or
\textquotedblleft TV\textquotedblright\ algorithm, which approximates the
evolution driven by $\hat{H}_{\text{eff}}=T(\hat{p})+V_{\text{eff}}(\hat
{q};\psi)$ with the first-order accuracy in the time step $\Delta t$.
Composing, in turn, the VT with TV algorithm, both with the time step $\Delta
t/2$, yields, depending on the order of composition, either the
\textquotedblleft VTV\textquotedblright\ or \textquotedblleft
TVT\textquotedblright\ second-order algorithm, which are analogues of the
Verlet algorithm\cite{Verlet:1967} for classical molecular dynamics and of the
split-operator algorithm\cite{Feit_Steiger:1982} for quantum dynamics. They
also generalize Faou and Lubich's algorithm\cite{Faou_Lubich:2006} for the
variational TGWD to the TGWD with a general effective
potential~(\ref{eq:V_eff}). Both VTV\ and TVT\ algorithms are symmetric and,
therefore, time-reversible.

Explicit geometric integrators of arbitrary even orders in $\Delta t$ are then
obtained by applying recursive\cite{Yoshida:1990,Suzuki:1990} or
nonrecursive\cite{Kahan_Li:1997,Sofroniou_Spaletta:2005} symmetric composition
schemes to the second-order
algorithms.\cite{book_Leimkuhler_Reich:2004,book_Hairer_Wanner:2006} For
details on how high-order integrators are generated from the elementary
algorithms for kinetic and potential propagations, see, e.g.,
Refs.~\onlinecite{Choi_Vanicek:2019,Roulet_Vanicek:2019,Roulet_Vanicek:2021a}.
Because this procedure is general, one can use for the splitting and
composition the same
algorithm\cite{Choi_Vanicek:2019,Roulet_Vanicek:2019,Roulet_Vanicek:2021a} as
for any other classical or quantum dynamics; below we only need to describe
exact solutions for the kinetic and potential propagations.

The algorithm presented below suggests how to write a single computer program
that can evaluate all methods from Secs.~\ref{sec:TG_methods} and
\ref{sec:FG_methods}, regardless whether they were derived using the
variational principle, Hamiltonian approach, or, more generally, perspective
of the nonlinear TDSE. One simply invokes the potential propagation with
different coefficients $V_{0}$, $V_{1}$, and $V_{2}$.

\subsection{Kinetic propagation}

Equations of motion for the kinetic propagation are obtained by considering
only the kinetic term in the effective Hamiltonian, i.e., by setting $\hat
{H}_{\text{eff}}=T(\hat{p})$, which results in a problem equivalent to solving
the propagation of a free-particle Gaussian wavepacket. Setting $V_{0}%
=V_{1}=V_{2}=0$ in Eqs.~(\ref{eq:q_dot})-(\ref{eq:gamma_dot}) yields the
system%
\begin{align}
\dot{q}_{t}  &  =m^{-1}\cdot p_{t},\label{eq:q_dot_T_prop}\\
\dot{p}_{t}  &  =0,\label{eq:p_dot_T_prop}\\
\dot{A}_{t}  &  =-A_{t}\cdot m^{-1}\cdot A_{t},\label{eq:A_dot_T_prop}\\
\dot{\gamma}_{t}  &  =T(p_{t})+(i\hbar/2)\operatorname*{Tr}\left(  m^{-1}\cdot
A_{t}\right)  \label{eq:gamma_dot_T_prop}%
\end{align}
of ordinary differential equations for parameters $\Lambda_{t}:=(q_{t}%
,p_{t},A_{t},\gamma_{t})$ whose exact solution is the flow $\Lambda_{t}%
=\Phi_{T}(\Lambda_{0},t)$ given explicitly by%
\begin{align}
q_{t}  &  =q_{0}+tm^{-1}\cdot p_{0},\label{eq:q_t_T_prop}\\
p_{t}  &  =p_{0},\label{eq:p_t_T_prop}\\
A_{t}  &  =\left(  A_{0}^{-1}+tm^{-1}\right)  ^{-1}=A_{0}\cdot\left(
\operatorname{Id}_{D}+tm^{-1}\cdot A_{0}\right)  ^{-1}\nonumber\\
&  =\left(  \operatorname{Id}_{D}+tA_{0}\cdot m^{-1}\right)  ^{-1}\cdot
A_{0},\label{eq:A_t_T_prop}\\
\gamma_{t}  &  =\gamma_{0}+tT(p_{0})+\frac{i\hbar}{2}\ln\det\left(
\operatorname{Id}_{D}+tm^{-1}\cdot A_{0}\right)  . \label{eq:gamma_t_T_prop}%
\end{align}
In the last equation, the continuity of $\gamma_{t}$ and, therefore,
continuity of $\psi(t)$, is guaranteed for sufficiently small time steps $t$
if one takes the principal branch of the logarithm---the branch on which the
imaginary part of the logarithm lies in the interval $(-\pi,\pi)$%
.\cite{Begusic_Vanicek:2022} The first of the three alternatives for
evaluating $A_{t}$ behaves better numerically despite requiring two instead of
one matrix inverse at each step ($m^{-1}$ can be precomputed). In the FGWD,
Eq.~(\ref{eq:A_dot_T_prop}) is replaced with $\dot{A}_{t}=0$,
Eq.~(\ref{eq:A_t_T_prop}) with $A_{t}=A_{0}$, and Eq.~(\ref{eq:gamma_t_T_prop}%
) with%
\begin{equation}
\gamma_{t}=\gamma_{0}+t[T(p_{0})-(\hbar/2)\operatorname{Tr}\left(  m^{-1}%
\cdot\mathcal{B}\right)  ].
\end{equation}

Because $T(\hat{p})$ is a special case of a linear Hamiltonian operator
$\hat{H}$, kinetic propagation conserves the scalar product, norm, and
symplectic structure. Kinetic propagation is also time-reversible because
$\Lambda_{0}=\Phi_{T}(\Lambda_{t},-t)$, which follows by inverting
Eqs.~(\ref{eq:q_t_T_prop})--(\ref{eq:gamma_t_T_prop}) explicitly:%
\begin{align}
q_{0}  &  =q_{t}-tm^{-1}\cdot p_{t},\label{eq:q_0_inv_T_prop}\\
p_{0}  &  =p_{t},\label{eq:p_0_inv_T_prop}\\
A_{0}  &  =\left(  A_{t}^{-1}-tm^{-1}\right)  ^{-1}, \label{eq:A_0_inv_T_prop}%
\\
\gamma_{0}  &  =\gamma_{t}-tT(p_{t})+\frac{i\hbar}{2}\ln\det\left(
\operatorname{Id}_{D}-tm^{-1}\cdot A_{t}\right)  .
\label{eq:gamma_0_inv_T_prop}%
\end{align}
In FGWD, Eqs.~(\ref{eq:A_0_inv_T_prop}) and (\ref{eq:gamma_0_inv_T_prop})
become $A_{0}=A_{t}$ and%
\begin{equation}
\gamma_{0}=\gamma_{t}+t[T(p_{t})-(\hbar/2)\operatorname{Tr}\left(  m^{-1}%
\cdot\mathcal{B}\right)  ].
\end{equation}
Equations~(\ref{eq:q_t_T_prop})--(\ref{eq:gamma_t_T_prop}) and
(\ref{eq:gamma_0_inv_T_prop}) for the forward and backward propagations are
derived in Appendix~\ref{app:kinetic_propagation_derivation}.

\subsection{\label{subsec:V_prop}Potential propagation}

Equations of motion for the potential propagation are obtained by considering
only the potential energy term in the effective Hamiltonian, i.e., $\hat
{H}_{\text{eff}}(\psi)=V_{\text{eff}}(\hat{q};\psi)$. Setting $T(p)=0$ is
equivalent to taking the limit $m\rightarrow\infty$ (or $m^{-1}\rightarrow0$)
in Eqs.~(\ref{eq:q_dot})-(\ref{eq:gamma_dot}) and yields the system%
\begin{align}
\dot{q}_{t}  &  =0,\label{eq:q_dot_V_prop}\\
\dot{p}_{t}  &  =-V_{1},\label{eq:p_dot_V_prop}\\
\dot{A}_{t}  &  =-V_{2},\label{eq:A_dot_V_prop}\\
\dot{\gamma}_{t}  &  =-V_{0}. \label{eq:gamma_dot_V_prop}%
\end{align}
This system can be solved analytically if the coefficients $V_{0}$, $V_{1}$,
$V_{2}$ depend on the state $\psi(t)$ only via $q_{t}$ and $\operatorname{Im}%
A_{t}$ but are independent of $p_{t}$, $\operatorname{Re}A_{t}$, and
$\gamma_{t}$. This assumption, which holds for all approximations from
Secs.~\ref{sec:TG_methods} and \ref{sec:FG_methods}, results in a trivial
solution%
\begin{align}
q_{t}  &  =q_{0},\label{eq:q_t_V_prop}\\
p_{t}  &  =p_{0}-tV_{1}(q_{0},\operatorname{Im}A_{0}),\label{eq:p_t_V_prop}\\
A_{t}  &  =A_{0}-tV_{2}(q_{0},\operatorname{Im}A_{0}),\label{eq:A_t_V_prop}\\
\gamma_{t}  &  =\gamma_{0}-tV_{0}(q_{0},\operatorname{Im}A_{0}).
\label{eq:gamma_t_V_prop}%
\end{align}
In FGWD, Eqs.~(\ref{eq:A_dot_V_prop}) and (\ref{eq:A_t_V_prop}) are replaced
with $\dot{A}_{t}=0$ and $A_{t}=A_{0}$.

Equation (\ref{eq:q_t_V_prop}) follows immediately from
Eq.~(\ref{eq:q_dot_V_prop}). Because $V_{2}$ is real,
Eq.~(\ref{eq:A_dot_V_prop}) implies that $d\operatorname{Im}A_{t}/dt=0$ and,
therefore, $\operatorname{Im}A_{t}=\operatorname{Im}A_{0}$. As a consequence,
if the coefficients $V_{j}\equiv V_{j}(q_{t},\operatorname{Im}A_{t})$ depend
only on $q_{t}$ and $\operatorname{Im}A_{t}$, then $V_{j}$ remain unchanged
during the potential propagation and Eqs.~(\ref{eq:p_dot_V_prop}%
)--(\ref{eq:gamma_dot_V_prop}) can be solved separately to yield
Eqs.~(\ref{eq:p_t_V_prop})--(\ref{eq:gamma_t_V_prop}), respectively.

The assumption is obviously satisfied in the global harmonic, local harmonic,
and single-Hessian TGWD, for which the coefficients of $V_{\text{eff}}$ depend
on $\psi(t)$ only via $q_{t}$. In the variational methods, coefficients of the
effective potential depend on the expectation values
\begin{equation}
\langle\hat{V}^{(j)}\rangle=\int V^{(j)}(q_{t}+x)\rho(x)d^{D}x,
\label{eq:exp_V_j}%
\end{equation}
which depend on both $q_{t}$ and $\operatorname{Im}A_{t}$ but on no other
parameters of $\psi(t)$ because the density $\rho(x)=|\psi(x)|^{2}$ depends on
$\psi(t)$ only via $\operatorname{Im}A_{t}$ [see Eq.~(\ref{eq:psi_squared})].

Potential propagation conserves the norm of $\psi(t)$ because $V_{\text{eff}%
}(\hat{q};\psi)$ is a special case of $\hat{H}_{\text{eff}}$. Potential
propagation is also time-reversible because $\Lambda_{0}=\Phi_{V_{\text{eff}}%
}(\Lambda_{t},-t)$, which, in turn, follows by inverting
Eqs.~(\ref{eq:q_t_V_prop})--(\ref{eq:gamma_t_V_prop}):%
\begin{align}
q_{0}  &  =q_{t},\label{eq:q_0_inv_V_prop}\\
p_{0}  &  =p_{t}+tV_{1}(q_{t},\operatorname{Im}A_{t}%
),\label{eq:p_0_inv_V_prop}\\
A_{0}  &  =A_{t}+tV_{2}(q_{t},\operatorname{Im}A_{t}%
),\label{eq:A_0_inv_V_prop}\\
\gamma_{0}  &  =\gamma_{t}+tV_{0}(q_{t},\operatorname{Im}A_{t}).
\label{eq:gamma_0_inv_V_prop}%
\end{align}
In FGWD, Eq.~(\ref{eq:A_0_inv_V_prop}) is replaced with $A_{0}=A_{t}$.

In general, propagation with $V_{\text{eff}}$ does not conserve symplectic
structure: potential propagation in neither the local harmonic nor local
quartic variational TGWD is symplectic.\cite{Ohsawa_Leok:2013} In contrast,
potential propagations in all other presented examples of the TGWD, i.e., the
variational, single-Hessian, global harmonic, local cubic variational, and
single-quartic variational TGWD are symplectic.

\subsection{Geometric properties of integrators}

If the exact solution of the nonlinear TDSE has a certain geometric property
for any Hamiltonian $\hat{H}_{\text{eff}}=T(\hat{p})+V_{\text{eff}}(\hat
{q};\psi)$, then both the kinetic and potential steps share the same property
because they can be thought of as exact solutions of nonlinear TDSEs with
effective Hamiltonians $\hat{H}_{\text{eff}}=T(\hat{p})$ and $\hat
{H}_{\text{eff}}(\psi)=V_{\text{eff}}(\hat{q};\psi)$. This implication
justifies the conservation of norm by kinetic and potential propagations.
Although the implication is also true for the exact and effective energies, it
is \textquotedblleft useless\textquotedblright\ because the definitions of
these energies are different for the three effective Hamiltonians $\hat{T}$,
$\hat{V}_{\text{eff}}$, and $\hat{T}+\hat{V}_{\text{eff}}$.

If a geometric property preserved by kinetic and potential steps is also
preserved under composition of flows, it is preserved by the TV and VT
integrators as well as by their arbitrary compositions. This is again true for
the norm and useless for the exact and effective energies. It is also true for
symplecticity; as a consequence, if the elementary propagation with
$V_{\text{eff}}$ is symplectic, then so is an arbitrary integrator based on
composing VT and TV steps. Time reversibility requires that the composition of
time-reversible maps be symmetric. VT\ and TV integrators are not reversible,
whereas TVT\ and VTV integrators and their symmetric compositions are reversible.

In summary, all integrators obtained by symmetric compositions of the VTV and
TVT algorithms are norm-conserving and time-reversible. They are also
symplectic \emph{if} the exact solution of the nonlinear TDSE itself is
symplectic. Due to the splitting, however, for a given finite time step
$\Delta t$ they conserve neither the exact nor the effective energy even if
the exact solution of the nonlinear TDSE does. [This happens for the exact
energy, conserved by the variational TGWD and variational FGWD. The numerical
integrators do conserve the exact energy, but only approximately, with an
error $O(\Delta t^{M})$, where the order $M$ is greater or equal to the order
of the method.] General proofs of these statements can be found in Refs.~\onlinecite{book_Leimkuhler_Reich:2004,book_Hairer_Wanner:2006,Choi_Vanicek:2019,Roulet_Vanicek:2019,Roulet_Vanicek:2021}.

\subsection{Geometric properties of Gaussian wavepacket dynamics}

Typically, one first demonstrates a geometric property of an approximation,
before analyzing the preservation of this property by a numerical integrator.
This is how we have treated norm conservation. However, sometimes it is easier
to go \textquotedblleft the other way.\textquotedblright\ If an integrator is
\textquotedblleft consistent,\textquotedblright\ i.e., at least first-order
accurate in the time step $\Delta t$, and preserves a given geometric property
during both kinetic or potential propagations, then considering the limit
$\Delta t\rightarrow0$ shows that the exact solution of the nonlinear TDSE has
the same geometric property. Application of this idea to the time-reversible
second-order TVT algorithm proves the time reversibility of the general
Gaussian wavepacket dynamics. Application of this idea to symplecticity shows
that if the potential propagation step with $V_{\text{eff}}$ is symplectic,
then so is the Gaussian wavepacket dynamics with this effective potential.
Whereas the symplecticity of the variational and local cubic variational TGWD
was demonstrated in Refs.~\onlinecite{Faou_Lubich:2006} and
\onlinecite{Ohsawa_Leok:2013}, a detailed analysis of symplecticity of the
TGWD with a general effective potential (\ref{eq:V_eff}) will be presented
elsewhere, because this analysis relies on nonelementary tools of symplectic
geometry.\cite{Vanicek:symplecticity}

\section{\label{sec:Hagedorn_parametrization}Hagedorn parametrization}

Hagedorn\cite{Hagedorn:1980} proposed an alternative parametrization of the
Gaussian wavepacket, in which the equations of motion and other properties
become simpler. Below we translate the preceding results from Heller's
parametrization $(q,p,A,\gamma)$ to Hagedorn's parametrization $(q,p,Q,P,S)$%
.\cite{Hagedorn:1998,book_Lubich:2008,Lasser_Lubich:2020} For derivations, see
Appendix~\ref{sec:Hagedorn_parametrization_derivation}.

\subsection{ Gaussian wavepacket}

As shown in Appendix~\ref{sec:Hagedorn_parametrization_derivation}, in
Hagedorn's parametrization, the Gaussian wavepacket~(\ref{eq:GWP}) can be
written as
\begin{align}
&  \psi(q,t)=\left(  \pi\hbar\right)  ^{-D/4}\left(  \det Q_{t}\right)
^{-1/2}\nonumber\\
&  \times\exp\left[  \frac{i}{\hbar}\left(  \frac{1}{2}x^{T}\cdot P_{t}\cdot
Q_{t}^{-1}\cdot x+p_{t}^{T}\cdot x+S_{t}\right)  \right]  ,
\label{eq:GWP_Hagedorn}%
\end{align}
where $q_{t}$ and $p_{t}$ are, as before, real $D$-vectors of position and
momentum, $Q_{t}$ and $P_{t}$ are complex $D\times D$ matrices, related to the
width matrix $A_{t}$, and $S_{t}$ is a real scalar generalizing classical
action and related to Heller's parameter $\gamma_{t}$. Matrices $Q_{t}$ and
$P_{t}$ have several remarkable properties, listed in
Appendix~\ref{sec:Hagedorn_parametrization_derivation}. In particular, the
often needed expressions for the position and momentum covariances assume the
symmetric and decoupled forms%
\begin{equation}
\operatorname{Cov}(\hat{q})=\frac{\hbar}{2}Q_{t}\cdot Q_{t}^{\dag}\text{
\ \ and \ \ }\operatorname{Cov}(\hat{p})=\frac{\hbar}{2}P_{t}\cdot P_{t}%
^{\dag}. \label{eq:Cov_q/p_Hagedorn}%
\end{equation}

\subsection{Equations of motion}

In Appendix~\ref{sec:Hagedorn_parametrization_derivation}, the nonlinear TDSE
(\ref{eq:tdse_q-rep}) is shown to be equivalent to a system of ordinary
differential equations for parameters $q_{t}$, $p_{t}$, $Q_{t}$, $P_{t}$, and
$S_{t}$. Whereas Eqs.~(\ref{eq:q_dot})--(\ref{eq:p_dot}) for $\dot{q}_{t}$ and
$\dot{p}_{t}$ remain unchanged, Eqs.~(\ref{eq:A_dot}) and (\ref{eq:gamma_dot})
for $\dot{A}_{t}$ and $\dot{\gamma}_{t}$ are replaced with three equations%
\begin{align}
\dot{Q}_{t}  &  =m^{-1}\cdot P_{t},\label{eq:Q_dot}\\
\dot{P}_{t}  &  =-V_{2}\cdot Q_{t},\label{eq:P_dot}\\
\dot{S}_{t}  &  =T(p_{t})-V_{0}. \label{eq:S_dot}%
\end{align}
A family of high-order geometric integrators for the numerical propagation can
be obtained, as in Heller's parametrization, by combining the concepts of
splitting into the sequence of kinetic and potential propagations, and of the
symmetric composition of the symmetric second-order TVT or VTV\ algorithm. We,
therefore, only need to derive expressions for the elementary kinetic and
potential propagations, which are generalizations of Faou and Lubich's
algorithm\cite{Faou_Lubich:2006,book_Lubich:2008} for the variational TGWD to
the generalized thawed Gaussian wavepacket dynamics expressed in
Eqs.~(\ref{eq:q_dot}), (\ref{eq:p_dot}), and (\ref{eq:Q_dot})-(\ref{eq:S_dot}).

\subsection{Kinetic propagation}

If the Hamiltonian consists only of the kinetic energy, $\hat{H}_{\text{eff}%
}=T(\hat{p})$, equations of motion (\ref{eq:Q_dot})--(\ref{eq:S_dot}) reduce
to%
\begin{align}
\dot{Q}_{t}  &  =m^{-1}\cdot P_{t},\label{eq:Q_dot_T_prop}\\
\dot{P}_{t}  &  =0,\label{eq:P_dot_T_prop}\\
\dot{S}_{t}  &  =T(p_{t}). \label{eq:S_dot_T_prop}%
\end{align}
Because the momentum $p_{t}$ is constant during the kinetic step, this system
has an analytical solution%
\begin{align}
Q_{t}  &  =Q_{0}+tm^{-1}\cdot P_{0},\label{eq:Q_t_T_prop}\\
P_{t}  &  =P_{0},\label{eq:P_t_T_prop}\\
S_{t}  &  =S_{0}+tT(p_{0}), \label{eq:S_t_T_prop}%
\end{align}
which is time-reversible [i.e., $\Lambda_{0}=\Phi_{T}(\Lambda_{t},-t)$ if
$\Lambda:=(q,p,Q,P,S)$ and $\Lambda_{t}=\Phi_{T}(\Lambda_{0},t)$] since the
inversion of Eqs.~(\ref{eq:Q_t_T_prop})--(\ref{eq:S_t_T_prop}) gives%
\begin{align}
Q_{0}  &  =Q_{t}-tm^{-1}\cdot P_{t},\\
P_{0}  &  =P_{t},\\
S_{0}  &  =S_{t}-tT(p_{t}).
\end{align}

\subsection{Potential propagation}

If $\hat{H}_{\text{eff}}(\psi)=V_{\text{eff}}(\hat{q};\psi)$,
Eqs.~(\ref{eq:Q_dot})--(\ref{eq:S_dot}) reduce to%
\begin{align}
\dot{Q}_{t}  &  =0,\label{eq:Q_dot_V_prop}\\
\dot{P}_{t}  &  =-V_{2}\cdot Q_{t},\label{eq:P_dot_V_prop}\\
\dot{S}_{t}  &  =-V_{0}. \label{eq:S_dot_V_prop}%
\end{align}
Since the position $q_{t}$ remains constant during the potential step, under
the assumption that the coefficients $V_{0}$, $V_{1}$, $V_{2}$ only depend on
$q_{t}$ and $Q_{t}$ but not on other Hagedorn parameters, this system has an
exact solution%
\begin{align}
Q_{t}  &  =Q_{0},\label{eq:Q_t_V_prop}\\
P_{t}  &  =P_{0}-tV_{2}(q_{0},Q_{0})\cdot Q_{0},\label{eq:P_t_V_prop}\\
S_{t}  &  =S_{0}-tV_{0}(q_{0},Q_{0}). \label{eq:S_t_V_prop}%
\end{align}
The assumption holds for all examples from Secs.~\ref{sec:TG_methods} and
\ref{sec:FG_methods}. In the global harmonic, local harmonic, and
single-Hessian TGWD, coefficients $V_{j}$ depend only on $q_{t}$; in the
variational methods, expected values $\langle\hat{V}^{(j)}\rangle$ and, hence,
coefficients $V_{j}$ depend only on $q_{t}$ and $Q_{t}$ because the density
$\rho(x)$ depends only on position covariance, which depends only on $Q_{t}$
[see Eqs.~(\ref{eq:exp_V_j}), (\ref{eq:psi_squared}) and
(\ref{eq:Cov_q/p_Hagedorn})]. Potential propagation is time-reversible, i.e.,
$\Lambda_{0}=\Phi_{V_{\text{eff}}}(\Lambda_{t},-t)$ if $\Lambda_{t}%
=\Phi_{V_{\text{eff}}}(\Lambda_{0},t)$, since the inversion of
Eqs.~(\ref{eq:Q_t_V_prop})--(\ref{eq:S_t_V_prop}) yields%
\begin{align}
Q_{0}  &  =Q_{t},\\
P_{0}  &  =P_{t}+tV_{2}(q_{t},Q_{t})\cdot Q_{t},\\
S_{0}  &  =S_{t}+tV_{0}(q_{t},Q_{t}).
\end{align}

\section{\label{sec:conclusion}Discussion and conclusion}

In conclusion, we have discussed the Gaussian wavepacket dynamics from the
perspective of a nonlinear Schr\"{o}dinger equation, which is complementary to
the variational\cite{Faou_Lubich:2006,book_Lubich:2008} and
symplectic\cite{Faou_Lubich:2006,Ohsawa_Leok:2013,Ohsawa:2015,Ohsawa:2015a}
perspectives. The more general state-dependent quadratic potential appearing
in the nonlinear TDSE describing Gaussian wavepacket dynamics guarantees norm
conservation and time reversibility but not always the conservation of energy,
effective energy, or symplectic structure. Depending on the choice of the
coefficients of this potential, one obtains a large family of both well-known
and new Gaussian wavepacket dynamics methods. Among the latter, the
single-quartic variational TGWD is promising because it is symplectic,
conserves the effective energy, and increases accuracy over the local cubic
variational TGWD without substantially increasing its cost.

The general form of presentation suggests how all single-trajectory Gaussian
wavepacket dynamics methods can be implemented in a single, universal computer
code, in which one only needs to modify the three coefficients of the
effective potential to obtain any one of the specific methods. Moreover, we
have described a single, universal high-order geometric integrator for
Gaussian wavepacket dynamics, which generalizes Faou and Lubich's integrator
for the variational TGWD.\cite{Faou_Lubich:2006}

Many but not all TGWD methods can be obtained by the variational and
symplectic approaches. If the variational principle is applied to the exact
potential ${V}$, the resulting method is symplectic and conserves
energy.\cite{Faou_Lubich:2006} Neither property is guaranteed if this
principle is applied to an approximate potential ${V}_{\text{{appr}}}$: local
harmonic approximation provides a counterexample. Remarkably, the symplectic
approach\cite{Ohsawa_Leok:2013} always conserves both the symplectic structure
and effective energy. If applied to the local harmonic approximation, the
symplectic method conserves the local harmonic energy, although the resulting
equations of motion are equivalent to those obtained by applying the
variational principle to the local cubic approximation. Heller's original
thawed Gaussian approximation therefore cannot be derived by the symplectic
approach. As opposed to both variational and symplectic approaches, the more
general nonlinear TDSE (\ref{eq:tdse_q-rep}) does not require a specific form
of the wavepacket: e.g., if it is applied to the local cubic approximation
directly, without invoking the variational principle, an initial Gaussian
wavepacket will quickly lose its Gaussian form.

Single-trajectory Gaussian wavepacket dynamics clearly cannot propagate
wavefunctions of more general forms, needed in many chemical physics
applications. This issue has been addressed partially in the extended thawed
Gaussian
approximation,\cite{Lee_Heller:1982,Patoz_Vanicek:2018,Prlj_Vanicek:2020,Begusic_Vanicek:2020}
which propagates a Gaussian multiplied with a linear polynomial, and thus can
describe electronic spectra beyond the Condon
approximation\cite{Patoz_Vanicek:2018,Prlj_Vanicek:2020} or rates of internal
conversion.\cite{Wenzel_Mitric:2023} The same issue is fully resolved by
Hagedorn wavepackets,\cite{Hagedorn:1980,Hagedorn:1998,Lasser_Lubich:2020}
which can propagate arbitrary wavefunctions. Indeed, Ohsawa generalized the
symplectic formulation to such non-Gaussian states.\cite{Ohsawa:2018} The
generalization of the present analysis to the dynamics of wavepackets of
arbitrary shapes is, therefore, also interesting and in progress. In
particular, any of the effective potentials described here will preserve the
form not only of Gaussian but also of Hagedorn wavepackets. Yet, even the
simple Gaussian wavepacket dynamics discussed here improves substantially
electronic spectra calculations over the standard global harmonic approaches,
which completely ignore the anharmonicity of the potential energy surface.
Among the different methods mentioned here, the
variational,\cite{Coalson_Karplus:1990} local cubic
variational,\cite{Pattanayak_Schieve:1994, Ohsawa_Leok:2013} and
single-quartic variational TGWD can even approximately capture tunneling and,
therefore, also deserve further attention.

\begin{acknowledgments}
The author thanks Roya Moghaddasi Fereidani for producing
Fig.~\ref{fig:GWP_and_V_eff}, Frank Grossmann, Fabian Kr\"{o}ninger, Caroline
Lasser, Christian Lubich, Tomoki Ohsawa, and Bill Poirier for discussions and
Tomislav Begu\v{s}i\'{c} for suggesting the proof in
Appendix~\ref{app:zero_ReA}. The author acknowledges the financial support
from the European Research Council (ERC) under the European Union's Horizon
2020 research and innovation program (grant agreement No. 683069 -- MOLEQULE)
as well as from the COST Action CA21101 -- Confined Molecular Systems: from a
new generation of materials to the stars (COSY) of the European Community.
\end{acknowledgments}

\section*{Author declarations}

\subsection*{Conflict of interest}

The author has no conflicts to disclose.

\section*{Data availability}

This study did not generate any data.

\appendix

\section{Properties of the nonlinear TDSE}

\subsection{\label{sec:d_psi_dot_phi_dt}Nonconservation of the inner product}

In Sec.~\ref{sec:nonlin_TDSE}, the nonconservation
(\ref{eq:time_dependence_of_sc_prod}) of the inner product by the nonlinear
TDSE follows from the relation%
\begin{align}
\langle\dot{\psi}|\phi\rangle &  =\langle\phi|\dot{\psi}\rangle^{\ast
}=-(i\hbar)^{-1}\langle\phi|\hat{H}_{\text{eff}}(\psi)\psi\rangle^{\ast
}\nonumber\\
&  =i\hbar^{-1}\langle\hat{H}_{\text{eff}}(\psi)\psi|\phi\rangle=i\hbar
^{-1}\langle\psi|\hat{H}_{\text{eff}}(\psi)\phi\rangle.
\label{eq:d_psi_dot_phi_dt}%
\end{align}

\subsection{\label{sec:time_dependence_of_exp_A}Evolution of the expected
value of a nonlinear operator}

To prove the relation (\ref{eq:nonlin_evol_exp_nonlin_A}), let us evaluate the
time derivative of a nonlinear observable $\langle\hat{A}(\psi)\rangle$:
\begin{align}
\frac{d}{dt}\langle\hat{A}(\psi)\rangle &  =\frac{d}{dt}\langle\psi|\hat
{A}(\psi)\psi\rangle\nonumber\\
&  =\langle\dot{\psi}|\hat{A}(\psi)\psi\rangle+\left\langle \psi\left\vert
\frac{d\hat{A}(\psi)}{dt}\psi\right.  \right\rangle +\langle\psi|\hat{A}%
(\psi)\dot{\psi}\rangle\nonumber\\
&  =i\hbar^{-1}\langle\psi|\hat{H}_{\text{eff}}(\psi)\hat{A}(\psi)\psi
\rangle+\langle d\hat{A}(\psi)/dt\rangle\nonumber\\
&  ~~~-i\hbar^{-1}\langle\psi|\hat{A}(\psi)\hat{H}_{\text{eff}}(\psi
)\psi\rangle\nonumber\\
&  =\langle d\hat{A}(\psi)/dt\rangle-i\hbar^{-1}\langle\lbrack\hat{A}%
(\psi),\hat{H}_{\text{eff}}(\psi)]\rangle,
\end{align}
In the third step of the derivation, we used the relation%
\begin{align}
\langle\dot{\psi}|\hat{A}(\psi)\psi\rangle &  =\langle\hat{A}(\psi)\psi
|\dot{\psi}\rangle^{\ast}=i\hbar^{-1}\langle\hat{A}(\psi)\psi|\hat
{H}_{\text{eff}}(\psi)\psi\rangle^{\ast}\nonumber\\
&  =i\hbar^{-1}\langle\hat{H}_{\text{eff}}(\psi)\psi|\hat{A}(\psi)\psi
\rangle\nonumber\\
&  =i\hbar^{-1}\langle\psi|\hat{H}_{\text{eff}}(\psi)\hat{A}(\psi)\psi\rangle,
\end{align}
which follows easily from the hermiticity of $\hat{H}_{\text{eff}}(\psi)$
viewed as a linear operator for a fixed state $\psi$.

\subsection{\label{sec:time_dependence_of_energy_in_separ_H}Time dependence of
energy, position, and momentum in a nonlinear TDSE with a separable
Hamiltonian}

For separable Hamiltonians, the commutator in the right hand side of
Eq.~(\ref{eq:dE_dt}) becomes%
\begin{equation}
\lbrack\hat{H},\hat{H}_{\text{eff}}(\psi)]=[\hat{V}-\hat{V}_{\text{eff}}%
(\psi),\hat{T}], \label{eq:comm_H_Heff}%
\end{equation}
where%
\begin{align}
2[\hat{V},\hat{T}]  &  =\hat{p}^{T}\cdot m^{-1}\cdot\lbrack\hat{V},\hat
{p}]+[\hat{V},\hat{p}^{T}]\cdot m^{-1}\cdot\hat{p}\nonumber\\
&  =i\hbar(\hat{p}^{T}\cdot m^{-1}\cdot\hat{V}^{\prime}+\hat{V}^{\prime
T}\cdot m^{-1}\cdot\hat{p}) \label{eq:comm_VT}%
\end{align}
and an analogous relation holds for $[\hat{V}_{\text{eff}}(\psi),\hat{T}]$.
Substituting Eqs.~(\ref{eq:comm_H_Heff}) and (\ref{eq:comm_VT}) into the
general Eq.~(\ref{eq:dE_dt}) for the time dependence of energy yields
Eq.~(\ref{eq:comm_H_Heff_1}).

Ehrenfest theorem [Eqs.~(\ref{eq:nonlinear_Ehrenfest_q}) and
(\ref{eq:nonlinear_Ehrenfest_p})] for the evolution of $q$ and $p$ follows
directly from the commutators%
\begin{align}
2[\hat{q}_{j},\hat{H}_{\text{eff}}]  &  =[\hat{q}_{j},\hat{p}^{T}\cdot
m^{-1}\cdot\hat{p}]\nonumber\\
&  =[\hat{q}_{j},\hat{p}^{T}]\cdot m^{-1}\cdot\hat{p}+\hat{p}^{T}\cdot
m^{-1}\cdot\lbrack\hat{q}_{j},\hat{p}]\nonumber\\
&  =2i\hbar(m^{-1}\cdot\hat{p})_{j},\label{eq:comm_q_and_H_eff}\\
\lbrack\hat{p},\hat{H}_{\text{eff}}]  &  =-i\hbar\hat{V}_{\text{eff}}^{\prime
}. \label{eq:comm_p_and_H_eff}%
\end{align}

\section{\label{sec:properties_of_GWP}Properties of the Gaussian wavepacket}

\subsection{Derivatives of functions of a vector or matrix}

We need several matrix relations, which can be found, e.g., in
Ref.~\onlinecite{Petersen_Pedersen:2012}. A derivative of a scalar function
$\psi(x)$ of a vector $x$ is defined by%
\begin{equation}
\lbrack\partial\psi(x)/\partial x]_{i}:=\partial\psi(x)/\partial x_{i}.
\label{eq:der_wrt_x}%
\end{equation}
Likewise, a derivative of a scalar function $\psi(A)$ of a general, not
necessarily symmetric, matrix $A$ is defined by%
\begin{equation}
\lbrack\partial\psi(A)/\partial A]_{ij}:=\partial\psi(A)/\partial A_{ij}.
\label{eq:der_wrt_A}%
\end{equation}
With these definitions, if $a$ is a vector, we get%
\begin{align}
\partial(a^{T}\cdot x)/\partial x  &  =a,\\
\partial\left(  x^{T}\cdot A\cdot x\right)  /\partial A  &  =x\otimes
x^{T},\label{eq:der_qform_wrt_A}\\
\partial\left(  x^{T}\cdot A\cdot x\right)  /\partial x  &  =(A+A^{T})\cdot x.
\label{eq:der_qform_wrt_x}%
\end{align}
If $B=B^{T}$ is a symmetric matrix, Eq.~(\ref{eq:der_qform_wrt_x}) reduces to%
\begin{equation}
\partial\left(  x^{T}\cdot B\cdot x\right)  /\partial x=2B\cdot x.
\label{eq:der_qform_wrt_x_symA}%
\end{equation}
If $x$ and $A$ depend on time $t$, then%
\begin{align}
d\psi(x)/dt  &  =(\partial\psi/\partial x_{i})\,\dot{x}_{i}=\left(
\partial\psi/\partial x\right)  ^{T}\cdot\dot{x},\label{eq:der_psi_wrt_x}\\
d\psi(A)/dt  &  =(\partial\psi/\partial A_{ij})\,\dot{A}_{ij}%
=\operatorname*{Tr}[\left(  \partial\psi/\partial A\right)  ^{T}\cdot\dot{A}].
\label{eq:der_psi_wrt_A}%
\end{align}
Divergence of a vector-valued function $v(A,x)=A\cdot x$ is%
\begin{equation}
\nabla^{T}\cdot v=\partial(A_{ij}x_{j})/\partial x_{i}=A_{ij}\delta
_{ji}=\operatorname{Tr}A. \label{eq:div_Ax}%
\end{equation}

\subsection{Derivatives of the Gaussian amplitude}

Derivatives of the Gaussian (\ref{eq:GWP}) with respect to the four parameters
are%
\begin{align}
\frac{\partial\psi}{\partial q_{t}}  &  =\left(  \frac{\partial x}{\partial
q_{t}}\right)  ^{T}\cdot\frac{\partial\psi}{\partial x}=-\frac{\partial\psi
}{\partial x}=-\frac{i}{\hbar}\xi\psi,\label{eq:dpsi_dq}\\
\partial\psi/\partial p_{t}  &  =i\hbar^{-1}x\psi,\label{eq:dpsi_dp}\\
\partial\psi/\partial A_{t}  &  =i\hbar^{-1}(1/2)x\otimes x^{T}\psi
,\label{eq:dpsi_dA}\\
\partial\psi/\partial\gamma_{t}  &  =i\hbar^{-1}\psi, \label{eq:dpsi_dgamma}%
\end{align}
where $\xi$ is given in Eq.~(\ref{eq:xi}) and
relation~(\ref{eq:der_qform_wrt_A}) was used to obtain Eq.~(\ref{eq:dpsi_dA}).
First and second derivatives of $\psi$ with respect to the coordinate $q$ (or
$x$), needed in $\langle q|\hat{T}|\psi\rangle$, are
\begin{align}
\nabla\psi &  =\partial\psi/\partial x=i\hbar^{-1}\xi\psi, \label{eq:grad_psi}%
\\
\nabla^{T}\cdot m^{-1}\cdot\nabla\psi &  =\frac{i}{\hbar}\left[  \left(
\nabla^{T}\cdot m^{-1}\cdot\xi\right)  \psi+(\nabla^{T}\psi)\cdot m^{-1}%
\cdot\xi\right] \nonumber\\
&  =\frac{i}{\hbar}\left[  \operatorname*{Tr}\left(  m^{-1}\cdot A_{t}\right)
+\frac{i}{\hbar}\xi^{T}\cdot m^{-1}\cdot\xi\right]  \psi, \label{eq:2T_psi}%
\end{align}
where Eqs.~(\ref{eq:div_Ax}) and (\ref{eq:dpsi_dq}) were used in the last step.

\subsection{Derivatives of the Gaussian density}

Probability density of the Gaussian (\ref{eq:GWP}) is%
\begin{equation}
\rho\left(  x\right)  :=\left\vert \psi\left(  x\right)  \right\vert ^{2}%
=\det\left(  2\pi\Sigma_{t}\right)  ^{-1/2}e^{-x^{T}\cdot\Sigma_{t}^{-1}\cdot
x/2}, \label{eq:psi_squared}%
\end{equation}
where $\Sigma_{t}=\operatorname*{Cov}(x)=\operatorname*{Cov}(q)$ is the
position covariance matrix (\ref{eq:Cov_q}), which is a real, symmetric,
positive definite, and invertible matrix because $\operatorname{Im}A_{t}$ is.

Derivatives of the Gaussian density (\ref{eq:psi_squared}) with respect to the
coordinate $q$ (or $x$) are
\begin{align}
\nabla\rho &  =\partial\rho/\partial x=-\Sigma_{t}^{-1}\cdot x\rho
,\label{eq:grad_psi_sq}\\
\nabla\otimes\nabla^{T}\rho &  =\left(  \Sigma_{t}^{-1}\cdot x\otimes
x^{T}\cdot\Sigma_{t}^{-1}-\Sigma_{t}^{-1}\right)  \rho. \label{eq:Hess_psi_sq}%
\end{align}
Because $\Sigma_{t}$ is invertible, the above equations for the gradient and
Hessian of the density imply that%
\begin{align}
x\rho &  =-\Sigma_{t}\cdot\nabla\rho,\label{eq:x_psi_sq}\\
x\otimes x^{T}\rho &  =\left(  \Sigma_{t}\cdot\nabla\otimes\nabla^{T}%
\cdot\Sigma_{t}+\Sigma_{t}\right)  \rho. \label{eq:x_sq_psi_sq}%
\end{align}

\subsection{\label{sec:covariances}Covariances}

Here we list explicit expressions for the frequently needed position and
momentum covariance matrices. Recall that the (generally complex)
cross-covariance matrix of Hermitian vector operators $\hat{A}$ and $\hat{B}$
is given by%
\begin{align}
\operatorname{Cov}(\hat{A},\hat{B})  &  :=\langle(\hat{A}-\langle\hat
{A}\rangle)\otimes(\hat{B}-\langle\hat{B}\rangle)^{T}\rangle\nonumber\\
&  =\langle\hat{A}\otimes\hat{B}^{T}\rangle-\langle\hat{A}\rangle
\otimes\langle\hat{B}\rangle^{T}. \label{eq:Cov_AB}%
\end{align}
and that one writes $\operatorname{Cov}(\hat{A})$ instead of
$\operatorname{Cov}(\hat{A},\hat{A})$ for the (always real)\ autocovariance of
$\hat{A}$. For $\hat{A}\neq\hat{B}$, a real cross-covariance matrix is
obtained by symmetrization:%
\begin{align}
\operatorname{Cov}_{R}(\hat{A},\hat{B})  &  =[\operatorname{Cov}(\hat{A}%
,\hat{B})+\operatorname{Cov}(\hat{B},\hat{A})^{T}]/2\nonumber\\
&  =\operatorname{Re}\operatorname{Cov}(\hat{A},\hat{B}). \label{eq:CovR_AB}%
\end{align}
The position, momentum, and position-momentum covariance matrices in a
Gaussian~(\ref{eq:GWP}) [or (\ref{eq:GWP_Hagedorn})] are%
\begin{align}
\operatorname{Cov}(\hat{q})  &  =(\hbar/2)\left(  \operatorname{Im}%
A_{t}\right)  ^{-1}=(\hbar/2)Q_{t}\cdot Q_{t}^{\dag},\label{eq:Cov_q}\\
\operatorname{Cov}(\hat{p})  &  =\frac{\hbar}{2}A_{t}\cdot\left(
\operatorname{Im}A_{t}\right)  ^{-1}\cdot A_{t}^{\ast}=\frac{\hbar}{2}%
P_{t}\cdot P_{t}^{\dag},\label{eq:Cov_p}\\
\operatorname{Cov}(\hat{q},\hat{p})  &  =\operatorname{Cov}(\hat{q})\cdot
A_{t}=(\hbar/2)\left(  \operatorname{Im}A_{t}\right)  ^{-1}\cdot
A_{t}\nonumber\\
&  =(\hbar/2)Q_{t}\cdot Q_{t}^{\dag}\cdot P_{t}\cdot Q_{t}^{-1}\nonumber\\
&  =\hbar(i+Q_{t}\cdot P_{t}^{\dag}/2),\label{eq:Cov_qp_Cov_q}\\
\operatorname{Cov}_{R}(\hat{q},\hat{p})  &  =(\hbar/2)\left(
\operatorname{Im}A_{t}\right)  ^{-1}\cdot\operatorname{Re}A_{t}\nonumber\\
&  =(\hbar/2)\operatorname{Re}(Q_{t}\cdot P_{t}^{\dag}), \label{eq:CovR_qp}%
\end{align}
where we have listed expressions in both Heller's and Hagedorn's
parametrizations. The first equality in the expression for $\operatorname{Cov}%
(\hat{q},\hat{p})$ holds because%
\begin{align}
\operatorname{Cov}(\hat{q},\hat{p})  &  =\langle(\hat{q}-q_{t})\otimes(\hat
{p}-p_{t})^{T}\rangle\nonumber\\
&  =\int\psi(x)^{\ast}x\otimes(\xi-p_{t})^{T}\psi(x)d^{D}x\nonumber\\
&  =\langle x\otimes(A_{t}\cdot x)^{T}\rangle=\langle x\otimes x^{T}%
\rangle\cdot A_{t},
\end{align}
whereas the last equality is obtained by the substitution $Q_{t}^{\dag}\cdot
P_{t}=2iI_{D}+P_{t}^{\dag}\cdot Q_{t}$, which follows from
Eq.~(\ref{eq:comm_QP2}).

We frequently need explicit expressions for the expectation value $\langle
\hat{A}\otimes\hat{B}^{T}\rangle$. Since $\langle\hat{q}\rangle=q_{t}$ and
$\langle\hat{p}\rangle=p_{t}$, the general relation (\ref{eq:Cov_AB}) implies
that%
\begin{align}
\langle\hat{q}\otimes\hat{q}^{T}\rangle &  =q_{t}\otimes q_{t}^{T}%
+\operatorname{Cov}(\hat{q}),\label{eq:exp_q_qT}\\
\langle\hat{p}\otimes\hat{p}^{T}\rangle &  =p_{t}\otimes p_{t}^{T}%
+\operatorname{Cov}(\hat{p}),\label{eq:exp_p_pT}\\
\langle\hat{q}\otimes\hat{p}^{T}\rangle &  =q_{t}\otimes p_{t}^{T}%
+\operatorname{Cov}(\hat{q},\hat{p}). \label{eq:exp_q_pT}%
\end{align}

For a frozen Gaussian, $\dot{A}_{t}=0$ and $A_{t}=A_{0}=\operatorname{const}$.
If $A_{t}=A_{0}$ is purely imaginary, i.e., $A_{0}=i\operatorname{Im}%
A_{0}=i\mathcal{B}$, covariance expressions (\ref{eq:Cov_q}%
)--(\ref{eq:CovR_qp}) reduce to%
\begin{align}
\operatorname{Cov}(\hat{q})  &  =(i\hbar/2)A_{0}^{-1}=(\hbar/2)\mathcal{B}%
^{-1},\label{eq:Cov_q_FG}\\
\operatorname{Cov}(\hat{p})  &  =-(i\hbar/2)A_{0}=(\hbar/2)\mathcal{B}%
,\label{eq:Cov_p_FG}\\
\operatorname{Cov}(\hat{q},\hat{p})  &  =i\hbar/2,\label{eq:Cov_qp_FG}\\
\operatorname{Cov}_{R}(\hat{q},\hat{p})  &  =0. \label{eq:CovR_qp_FG}%
\end{align}
Because the position covariance appears frequently in the text, we use a
shorthand notation $\Sigma_{t}:=\operatorname{Cov}(\hat{q}).$

\subsection{\label{sec:E_dot_in_TGWD}Time dependence of energy in the TGWD}

To gain further insight into the time dependence of energy, let us express the
expected value in Eq.~(\ref{eq:dE_dt_2}) for $\dot{E}$ in terms of differences
$\langle\hat{V}^{\prime}\rangle-V_{1}$ and $\langle\hat{V}^{\prime\prime
}\rangle-V_{2}$; to simplify the derivation, we employ notation $\hat{y}%
:=\hat{p}-p_{t}$ for the displaced momentum operator:%
\begin{align}
&  \langle(\hat{V}^{\prime}-V_{1}-V_{2}\cdot\hat{x})\otimes(p_{t}+\hat{y}%
)^{T}\rangle\nonumber\\
&  =\langle(\hat{V}^{\prime}-V_{1})\otimes(p_{t}+\hat{y})^{T}\rangle
-V_{2}\cdot\langle\hat{x}\otimes(p_{t}+\hat{y})^{T}\rangle\nonumber\\
&  =(\langle\hat{V}^{\prime}\rangle-V_{1})\otimes p_{t}^{T}+\langle\hat
{V}^{\prime}\otimes\hat{y}^{T}\rangle-V_{2}\cdot\langle\hat{x}\otimes\hat
{y}^{T}\rangle\nonumber\\
&  =(\langle\hat{V}^{\prime}\rangle-V_{1})\otimes p_{t}^{T}+\langle\hat
{V}^{\prime\prime}\rangle\cdot\operatorname{Cov}(\hat{q},\hat{p})-V_{2}%
\cdot\operatorname{Cov}(\hat{q},\hat{p})\nonumber\\
&  =(\langle\hat{V}^{\prime}\rangle-V_{1})\otimes p_{t}^{T}+(\langle\hat
{V}^{\prime\prime}\rangle-V_{2})\cdot\operatorname{Cov}(\hat{q},\hat{p}).
\label{eq:dE_dt_auxiliary}%
\end{align}
In the second step, we used $\left\langle \hat{x}\right\rangle =\left\langle
\hat{y}\right\rangle =0$ and, in the third step, invoked the definition
(\ref{eq:Cov_qp_Cov_q}) of the position-momentum covariance
$\operatorname{Cov}(\hat{q},\hat{p})$ and a useful relation%
\begin{equation}
\langle\hat{V}^{\prime}\otimes\left(  \hat{p}-p_{t}\right)  ^{T}%
\rangle=\langle\hat{V}^{\prime\prime}\rangle\cdot\operatorname{Cov}(\hat
{q},\hat{p}), \label{eq:exp_V_prime_p-pt}%
\end{equation}
proven in Appendix~\ref{sec:proof_of_V_prime_p-pt}. Substitution of
Eq.~(\ref{eq:dE_dt_auxiliary}) in Eq.~(\ref{eq:dE_dt_2}) for $\dot{E}$ yields
the final expression~(\ref{eq:dE_dt_3}) for the time dependence of energy.

\subsection{\label{sec:proof_of_V_prime_p-pt}Proof of
Eq.~(\ref{eq:exp_V_prime_p-pt}): $\langle\hat{V}^{\prime}\otimes\left(
\hat{p}-p_{t}\right)  ^{T}\rangle=\langle\hat{V}^{\prime\prime}\rangle
\cdot\operatorname{Cov}(\hat{q},\hat{p})$}

Equation (\ref{eq:exp_V_prime_p-pt}) follows immediately from the relations%
\begin{align}
\langle\hat{V}^{\prime}\otimes\left(  \hat{p}-p_{t}\right)  ^{T}\rangle &
=\langle\hat{V}^{\prime}\otimes\hat{x}^{T}\rangle\cdot A_{t}%
,\label{eq:exp_V_prime_p-pt_1}\\
\langle\hat{V}^{\prime}\otimes x^{T}\rangle &  =\langle\hat{V}^{\prime\prime
}\rangle\cdot\operatorname{Cov}(\hat{q}), \label{eq:exp_V_prime_x}%
\end{align}
and Eq.~(\ref{eq:Cov_qp_Cov_q}) for the position-momentum covariance. To prove
Eq.~(\ref{eq:exp_V_prime_p-pt_1}), we note that%
\begin{align}
&  \langle\hat{V}^{\prime}\otimes\left(  \hat{p}-p_{t}\right)  ^{T}%
\rangle\nonumber\\
&  =\int\psi(q,t)^{\ast}V^{\prime}(q)\otimes(-i\hbar\nabla-p_{t})^{T}%
\psi(q,t)d^{D}q\nonumber\\
&  =\int\psi(x,t)^{\ast}V^{\prime}(q_{t}+x)\otimes\left(  A_{t}\cdot x\right)
^{T}\psi(x,t)d^{D}x\nonumber\\
&  =\langle\hat{V}^{\prime}\otimes\hat{x}^{T}\rangle\cdot A_{t}.
\end{align}
Equation (\ref{eq:exp_V_prime_x}) follows from Eq.~(\ref{eq:x_psi_sq}) and
by-parts integration:%
\begin{align}
\langle\hat{V}^{\prime}\otimes\hat{x}^{T}\rangle &  =\int V^{\prime}%
(q_{t}+x)\otimes x^{T}\rho(x)d^{D}x\nonumber\\
&  =-\int\nabla V(q_{t}+x)\otimes\nabla^{T}\rho(x)d^{D}x\cdot
\operatorname{Cov}(\hat{q})\nonumber\\
&  =\int\rho(x)\nabla\otimes\nabla^{T}V(q_{t}+x)d^{D}x\cdot\operatorname{Cov}%
(\hat{q})\nonumber\\
&  =\left\langle \nabla\otimes\nabla^{T}V\right\rangle \cdot\operatorname{Cov}%
(\hat{q}).
\end{align}

\subsection{\label{sec:Hagedorn_parametrization_derivation}Derivation of the
Gaussian wavepacket's form and of the equations of motion in Hagedorn's
parametrization}

Here we justify Hagedorn's form~(\ref{eq:GWP_Hagedorn}) of the Gaussian
wavepacket as well as Eqs.~(\ref{eq:Q_dot})--(\ref{eq:S_dot}) of motion for
Hagedorn's parameters $Q_{t}$, $P_{t}$, and $S_{t}$.

Every complex symmetric matrix $A$ with a positive definite imaginary part can
be factorized as\cite{Lasser_Lubich:2020}
\begin{equation}
A=P\cdot Q^{-1}, \label{eq:A_PQ}%
\end{equation}
where $Q$ and $P$ are complex invertible matrices such that the real
$2D\times2D$ matrix%
\begin{equation}
Y:=%
\begin{pmatrix}
\operatorname{Re}Q & \operatorname{Im}Q\\
\operatorname{Re}P & \operatorname{Im}P
\end{pmatrix}
\end{equation}
is symplectic, i.e.,
\begin{equation}
Y^{T}\cdot J\cdot Y=J,
\end{equation}
where $J$ is the standard symplectic matrix%
\begin{equation}
J=%
\begin{pmatrix}
0 & -\operatorname{Id}_{D}\\
\operatorname{Id}_{D} & 0
\end{pmatrix}
\label{eq:J}%
\end{equation}
This condition is equivalent\cite{Lasser_Lubich:2020} to the requirement that
the matrices $Q$ and $P$ satisfy the relations%
\begin{align}
Q^{T}\cdot P-P^{T}\cdot Q  &  =0,\label{eq:comm_QP}\\
Q^{\dag}\cdot P-P^{\dag}\cdot Q  &  =2i\operatorname{Id}_{D}.
\label{eq:comm_QP2}%
\end{align}
Imaginary part of $A$ can be computed from $Q$ as\cite{Lasser_Lubich:2020}%
\begin{equation}
\operatorname{Im}A=(Q\cdot Q^{\dag})^{-1}. \label{eq:ImA_QQdag}%
\end{equation}

If one imposes the equation of motion
\begin{equation}
\dot{Q}_{t}=m^{-1}\cdot P_{t} \label{eq:Q_dot_imposed}%
\end{equation}
for $Q_{t}$, then $P_{t}$ must satisfy the equation
\begin{align}
\dot{P}_{t}  &  =\dot{A}_{t}\cdot Q_{t}+A_{t}\cdot\dot{Q}_{t}=-A_{t}\cdot
m^{-1}\cdot A_{t}\cdot Q_{t}\nonumber\\
&  \ \ \ -V_{2}\cdot Q_{t}+A_{t}\cdot m^{-1}\cdot P_{t}=-V_{2}\cdot Q_{t}.
\label{eq:P_dot_derived}%
\end{align}
In the derivation we used the Leibniz rule, Eq.~(\ref{eq:A_dot}) for $\dot
{A}_{t}$, and factorization (\ref{eq:A_PQ}) of $A_{t}$. It can be
shown\cite{Lasser_Lubich:2020} that if parameters $Q_{t}$ and $P_{t}$ are
propagated with Eqs.~(\ref{eq:Q_dot_imposed}) and (\ref{eq:P_dot_derived}) and
satisfy relations (\ref{eq:comm_QP})-(\ref{eq:ImA_QQdag}) at time zero, they
satisfy those relations for all times.

Equation~(\ref{eq:gamma_dot}) for $\dot{\gamma}_{t}$ can be rewritten as%
\begin{equation}
\dot{\gamma}_{t}=\dot{S}_{t}+\frac{i\hbar}{2}\frac{d}{dt}\ln\det Q_{t},
\label{eq:gamma_dot_QP_derivation}%
\end{equation}
where we have defined a generalized real action $S_{t}$ by%
\begin{equation}
\dot{S}_{t}=T(p_{t})-V_{0}. \label{eq:S_dot_derivation}%
\end{equation}
The second term in Eq.~(\ref{eq:gamma_dot_QP_derivation}) follows from
Eq.~(\ref{eq:gamma_dot}) since%
\begin{align}
\operatorname{Tr}\left(  m^{-1}\cdot A_{t}\right)   &  =\operatorname{Tr}%
\left(  m^{-1}\cdot P_{t}\cdot Q_{t}^{-1}\right)  =\operatorname{Tr}(\dot
{Q}_{t}\cdot Q_{t}^{-1})\nonumber\\
&  =\left(  \det Q_{t}\right)  ^{-1}\cdot d(\det Q_{t})/dt.
\end{align}
Here, we used the factorization (\ref{eq:A_PQ}) of $A_{t}$, equation of motion
(\ref{eq:Q_dot_imposed}) for $Q_{t}$, and the formula for the derivative of a
determinant. By integrating Eq.~(\ref{eq:gamma_dot_QP_derivation}) for
$\dot{\gamma}_{t}$, we can rewrite the wavepacket (\ref{eq:GWP}) at time $t$
as%
\begin{align}
&  \psi(q,t)=\left(  \det Q_{0}/\det Q_{t}\right)  ^{1/2}\nonumber\\
&  \times\exp\left[  \frac{i}{\hbar}\left(  \frac{1}{2}x^{T}\cdot P_{t}\cdot
Q_{t}^{-1}\cdot x+p_{t}^{T}\cdot x+S_{t}-S_{0}+\gamma_{0}\right)  \right]  .
\end{align}
Recall that the initial wavepacket $\psi(q,0)$ is normalized if $\gamma_{0}$
satisfies Eq. (\ref{eq:initial_GWP_normalization}). Setting the initial value
of $\gamma_{t}$ to $\gamma_{0}=S_{0}+i\operatorname{Im}\gamma_{0}$, where
\begin{align*}
\operatorname{Im}\gamma_{0}  &  =-\hbar\ln\det(\operatorname{Im}A_{0}/\pi
\hbar)^{1/4}=\hbar\ln\det(\pi\hbar Q_{0}\cdot Q_{0}^{\dag})^{1/4}\\
&  =\hbar\ln[(\pi\hbar)^{D}|\det Q_{0}|^{2}]^{1/4}=\frac{\hbar}{2}\ln
[(\pi\hbar)^{D/2}|\det Q_{0}|],
\end{align*}
we obtain a simple formula for the wavepacket parametrized by $Q_{t}$, $P_{t}%
$, and $S_{t}$ instead of $A_{t}$ and $\gamma_{t}$:%
\begin{align}
&  \psi(x,t)=\left(  \pi\hbar\right)  ^{-D/4}\left(  \frac{\det Q_{0}}{\det
Q_{t}|\det Q_{0}|}\right)  ^{1/2}\nonumber\\
&  \times\exp\left[  \frac{i}{\hbar}\left(  \frac{1}{2}x^{T}\cdot P_{t}\cdot
Q_{t}^{-1}\cdot x+p_{t}^{T}\cdot x+S_{t}\right)  \right]  .
\end{align}
The prefactor can be further simplified with another choice of the initial
value of $\gamma_{t}$, namely%
\begin{equation}
\gamma_{0}=S_{0}+\hbar\varphi_{0}+i\operatorname{Im}\gamma_{0},
\end{equation}
where $\varphi_{0}=-\frac{1}{2}\arg\det Q_{0}$, for which the wavepacket
assumes a simple Hagedorn form~(\ref{eq:GWP_Hagedorn}), valid at all times.

\section{\label{sec:cons_E_norm_by_DFVP}Dirac--Frenkel variational principle,
nonlinear TDSE, and conservation of energy and norm}

One often seeks an approximate solution of the TDSE only within a certain
subset $M$ of the full Hilbert space $\mathcal{H}$. If one seeks a solution
among states $\phi(t;\theta_{1},\,\ldots,\theta_{N})$ $\in\mathcal{H}$ that
depend on $N$ parameters $\theta_{j}$, $j=1,\ldots,N$, the Dirac-Frenkel
variational principle states that the optimal solution satisfies the equation%
\begin{equation}
\langle\delta\phi|[i\hbar(d/dt)-\hat{H}]|\phi(t)\rangle=0, \label{eq:TDVP}%
\end{equation}
where $\delta\phi$ is an arbitrary variation of the solution, i.e., an
infinitesimal change of $\phi$ such that $\phi+\delta\phi$ is still in the
approximation manifold $M$. [More precisely, $\delta\phi$ is an arbitrary
\textquotedblleft tangent vector\textquotedblright\ to the manifold $M$ at the
point $\phi(t)$.] The variation $\delta\phi$ can be expressed in terms of the
variations of its $N$ parameters as%
\begin{equation}
\delta\phi=\sum_{j=1}^{N}\frac{\partial\phi}{\partial\theta_{j}}\delta
\theta_{j}.
\end{equation}
Because the variations of parameters are independent, the Dirac-Frenkel
principle (\ref{eq:TDVP})\ requires the following equation to be satisfied for
each of the parameters:%
\begin{equation}
0=\left\langle \frac{\partial\phi}{\partial\theta_{j}}\left\vert \left(
i\hbar\frac{d}{dt}-\hat{H}\right)  \right\vert \phi\right\rangle \text{, for
}j=1,\ldots,N. \label{eq:DFVP_theta_j}%
\end{equation}

This\ variational principle provides a rich class of nonlinear TDSEs because
Eq.~(\ref{eq:TDVP}) is equivalent\cite{Lasser_Lubich:2020} to%
\begin{equation}
i\hbar|\dot{\phi}(t)\rangle=\hat{P}(\phi(t))\hat{H}|\phi(t)\rangle,
\end{equation}
where $\hat{P}(\phi)$ is the projection on the tangent space of $M$ at the
state $\phi$. In other words, the solution $\phi(t)$ satisfies the nonlinear
TDSE (\ref{eq:nonlinear_tdse}) with an effective Hamiltonian%
\begin{equation}
\hat{H}_{\text{eff}}(\phi):=\hat{P}(\phi(t))\hat{H}.
\end{equation}

Remarkably, solutions of the Dirac-Frenkel Eq.~(\ref{eq:TDVP}) preserve
several properties of the exact solution of the TDSE (\ref{eq:tdse}). In
particular, the energy is conserved along the variational solutions $\phi(t)$
satisfying Eq.~(\ref{eq:TDVP}) and the conservation of norm requires only a
weak additional assumption. The energy is conserved
because\cite{book_Lubich:2008}
\begin{align}
\frac{d}{dt}\langle\hat{H}\rangle &  =\frac{d}{dt}\langle\phi\left(  t\right)
|\hat{H}|\phi\left(  t\right)  \rangle=\langle\dot{\phi}|\hat{H}|\phi
\rangle+\langle\phi|\hat{H}|\dot{\phi}\rangle\nonumber\\
&  =2\operatorname{Re}\langle\dot{\phi}|\hat{H}|\phi\rangle=2\operatorname{Re}%
\langle\dot{\phi}|i\hbar\dot{\phi}\rangle\nonumber\\
&  =2\operatorname{Re}(i\hbar\Vert\dot{\phi}\Vert^{2})=0,
\end{align}
where, in the fourth step, we invoked the variational principle (\ref{eq:TDVP}%
) with $\delta\phi\propto\dot{\phi}$. If the manifold has the ray property
($\lambda\phi\in M$ for each complex number $\lambda$ and each $\phi\in M$),
then the norm remains constant because\cite{book_Lubich:2008}%
\begin{align}
\frac{d}{dt}\left\Vert \phi\left(  t\right)  \right\Vert ^{2}  &  =\frac
{d}{dt}\langle\phi\left(  t\right)  |\phi\left(  t\right)  \rangle=\langle
\dot{\phi}|\phi\rangle+\langle\phi|\dot{\phi}\rangle=2\operatorname{Re}%
\langle\phi|\dot{\phi}\rangle\nonumber\\
&  =2\hbar^{-1}\operatorname{Im}\langle\phi|i\hbar\dot{\phi}\rangle
=2\hbar^{-1}\operatorname{Im}\langle\phi|\hat{H}\phi\rangle\nonumber\\
&  =2\hbar^{-1}\operatorname{Im}\langle\hat{H}\rangle=0,
\end{align}
where, in the fifth step, we invoked the variational principle (\ref{eq:TDVP})
with $\delta\phi\propto\phi$. $\phi$ is in the tangent space because of the
ray property. If the ray property does not hold with given parameters, it will
hold if we augment\cite{Joubert-Doriol_Izmaylov:2018} the parameter set by a
prefactor $\lambda$ of the state $\phi$.

Note that the conservation of energy requires only the real part of
Eq.~(\ref{eq:TDVP}), which is sometimes referred to as the \emph{Lagrangian
variational principle }(or \emph{time-dependent variational principle}%
),\cite{Kramer_Saraceno:1981} whereas the conservation of norm requires only
the imaginary part of Eq.~(\ref{eq:TDVP}), which is sometimes referred to as
the \emph{McLachlan variational principle}.\cite{McLachlan:1964} For Gaussian
wavepackets, the three forms of the variational principle are
equivalent.\cite{Broeckhove_Lathouwers:1988,book_Lubich:2008,Lasser_Lubich:2020}%

\section{\label{sec:VGA}Variational Gaussian approximation}

The variational Gaussian approximation (or variational TGWD) follows from the
Dirac-Frenkel principle:

\textbf{Proposition (Variational TGWD}) The Dirac-Frenkel variational
principle (\ref{eq:TDVP}) for the TDSE (\ref{eq:tdse}) applied to the Gaussian
ansatz (\ref{eq:GWP}) yields the following equations of motion for the
Gaussian's parameters:
\begin{align}
\dot{q}_{t}  &  =m^{-1}\cdot p_{t},\label{eq:q_dot_VG}\\
\dot{p}_{t}  &  =-\langle\hat{V}^{\prime}\rangle,\label{eq:p_dot_VG}\\
\dot{A}_{t}  &  =-A_{t}\cdot m^{-1}\cdot A_{t}-\langle\hat{V}^{\prime\prime
}\rangle,\label{eq:A_dot_VG}\\
\dot{\gamma}_{t}  &  =T(p_{t})-\langle\hat{V}\rangle+(\hbar
/4)\operatorname*{Tr}[\langle\hat{V}^{\prime\prime}\rangle\cdot\left(
\operatorname{Im}A_{t}\right)  ^{-1}]\nonumber\\
&  ~~~+(i\hbar/2)\operatorname*{Tr}\left(  m^{-1}\cdot A_{t}\right)  .
\label{eq:gamma_dot_VG}%
\end{align}
These are equivalent to Eqs.~(\ref{eq:q_dot})--(\ref{eq:gamma_dot}) for the
TGWD satisfying the nonlinear TDSE (\ref{eq:tdse_q-rep}) with the effective
potential (\ref{eq:V_eff}) and coefficients~(\ref{eq:V_VGA}).

\emph{Proof}. Since the manifold of Gaussian wavepackets has the ray property,
the variational solution conserves the norm of the wavefunction (see
Appendix~\ref{sec:cons_E_norm_by_DFVP}).\cite{book_Lubich:2008} For the
Gaussian ansatz (\ref{eq:GWP}), this implies [see
Eq.~(\ref{eq:norm_squared_GWP})] that
\begin{equation}
\exp\left(  -\operatorname{Im}\gamma_{t}/\hbar\right)  =[\det
(\operatorname{Im}A_{t}/\pi\hbar)]^{1/4}%
\end{equation}
for all times $t$ if the initial norm is $1$. For Gaussian (\ref{eq:GWP}) with
density $\rho(x)$ [Eq.~(\ref{eq:psi_squared})], variational
equations~(\ref{eq:DFVP_theta_j}) for parameters $q_{t}$, $p_{t}$, $A_{t}$,
and $\gamma_{t}$ are, respectively,%
\begin{align}
0  &  =\int\left(  A_{t}\cdot x+p_{t}\right)  g(x)\rho\left(  x\right)
d^{D}x,\label{eq:DFVP_q}\\
0  &  =\int xg(x)\rho\left(  x\right)  d^{D}x,\label{eq:DFVP_p}\\
0  &  =\int x\otimes x^{T}g(x)\rho\left(  x\right)  d^{D}x,\label{eq:DFVP_A}\\
0  &  =\int g(x)\rho\left(  x\right)  d^{D}x, \label{eq:DFVP_gamma}%
\end{align}
where%
\begin{equation}
g(x):=f(x)-V(q_{t}+x) \label{eq:g}%
\end{equation}
is the difference between the quadratic polynomial (\ref{eq:quad_pol_f}) and
the potential energy $V\left(  q\right)  $. Because $\rho$ is normalized, the
variational equations (\ref{eq:DFVP_q})-(\ref{eq:DFVP_gamma}) can be expressed
as%
\begin{align}
0  &  =\left\langle \left(  A_{t}\cdot x+p_{t}\right)  g(x)\right\rangle
,\label{eq:DFVP_q_exp}\\
0  &  =\left\langle xg(x)\right\rangle ,\label{eq:DFVP_p_exp}\\
0  &  =\left\langle x\otimes x^{T}g(x)\right\rangle ,\label{eq:DFVP_A_exp}\\
0  &  =\left\langle g(x)\right\rangle . \label{eq:DFVP_gamma_exp}%
\end{align}
The first equation, which follows from the second and fourth equations, is
redundant. The last three equations, which are independent, are equivalent to
the system:%
\begin{align}
0  &  =\left\langle g(x)\right\rangle ,\label{eq:exp_g}\\
0  &  =\left\langle \nabla g(x)\right\rangle ,\label{eq:exp_grad_g}\\
0  &  =\left\langle \nabla\otimes\nabla^{T}g(x)\right\rangle .
\label{eq:exp_Hess_g}%
\end{align}
Equation~(\ref{eq:exp_grad_g}) is equivalent to Eq.~(\ref{eq:DFVP_p_exp})
because%
\begin{align}
\left\langle \nabla g(x)\right\rangle  &  =\int\rho\nabla g\,d^{D}x=-\int
g\nabla\rho\,d^{D}x\nonumber\\
&  =\int g\Sigma_{t}^{-1}\cdot x\rho\,d^{D}x=\Sigma_{t}^{-1}\cdot\left\langle
xg(x)\right\rangle .
\end{align}
Here, we integrated by parts and used relation (\ref{eq:grad_psi_sq}).
Likewise, Eqs.~(\ref{eq:DFVP_A_exp}) and (\ref{eq:DFVP_gamma_exp}) are
equivalent to Eqs.~(\ref{eq:exp_g}) and (\ref{eq:exp_Hess_g}) since%
\begin{align*}
\left\langle \nabla\otimes\nabla^{T}g(x)\right\rangle  &  =\int\rho
\nabla\otimes\nabla^{T}g\,d^{D}x=\int g\nabla\otimes\nabla^{T}\rho\,d^{D}x\\
&  =\int g\left(  \Sigma_{t}^{-1}\cdot x\otimes x^{T}\cdot\Sigma_{t}%
^{-1}-\Sigma_{t}^{-1}\right)  \rho\,d^{D}x\\
&  =\Sigma_{t}^{-1}\cdot\left\langle x\otimes x^{T}g(x)\right\rangle
\cdot\Sigma_{t}^{-1}-\Sigma_{t}^{-1}\left\langle g(x)\right\rangle ,
\end{align*}
where we integrated by parts twice and used Eq.~(\ref{eq:Hess_psi_sq}).

Recalling definition (\ref{eq:g}) of $g(x)$, we find that $g$, its gradient,
and its Hessian can be written explicitly as%
\begin{align}
g(x)  &  =x^{T}\cdot C_{2}\cdot x/2+x^{T}\cdot C_{1}+C_{0}-V(q_{t}%
+x),\label{eq:g_expl}\\
g^{\prime}(x)  &  =C_{2}\cdot x+C_{1}-V^{\prime}(q_{t}%
+x),\label{eq:grad_g_expl}\\
g^{\prime\prime}(x)  &  =C_{2}-V^{\prime\prime}(q_{t}+x).
\label{eq:Hess_g_expl}%
\end{align}
Substituting expressions (\ref{eq:g_expl})-(\ref{eq:Hess_g_expl}) for $g$,
$g^{\prime}$, and $g^{\prime\prime}$ into Eqs.~(\ref{eq:exp_g}%
)-(\ref{eq:exp_Hess_g}) for the expectation values yields%
\begin{align}
0  &  =\left\langle x^{T}\cdot C_{2}\cdot x\right\rangle /2+C_{1}^{T}%
\cdot\left\langle x\right\rangle +C_{0}-\langle\hat{V}\rangle\nonumber\\
&  =\operatorname*{Tr}\left(  C_{2}\cdot\Sigma_{t}\right)  /2+C_{0}%
-\langle\hat{V}\rangle,\label{eq:exp_g_expl}\\
0  &  =C_{2}\cdot\left\langle x\right\rangle +C_{1}-\langle\hat{V}^{\prime
}\rangle=C_{1}-\langle\hat{V}^{\prime}\rangle,\label{eq:exp_grad_g_expl}\\
0  &  =C_{2}-\langle\hat{V}^{\prime\prime}\rangle, \label{eq:exp_Hess_g_expl}%
\end{align}
where we have used relations $\langle x\rangle=0$ for the position and%
\begin{equation}
\left\langle x^{T}\cdot C_{2}\cdot x\right\rangle =\operatorname{Tr}\left(
C_{2}\cdot\left\langle x\otimes x^{T}\right\rangle \right)  =\operatorname{Tr}%
\left(  C_{2}\cdot\Sigma_{t}\right)
\end{equation}
for the expectation value of a quadratic form.

Equations (\ref{eq:exp_g_expl})-(\ref{eq:exp_Hess_g_expl}) have obvious
solutions%
\begin{align}
C_{0}  &  =\langle\hat{V}\rangle-\operatorname*{Tr}[\langle\hat{V}%
^{\prime\prime}\rangle\cdot\Sigma_{t}]/2,\label{eq:C0_VG}\\
C_{1}  &  =\langle\hat{V}^{\prime}\rangle,\label{eq:C1_VG}\\
C_{2}  &  =\langle\hat{V}^{\prime\prime}\rangle. \label{eq:C2_VG}%
\end{align}
In view of Propositions 1 and 2 from Sec.~\ref{subsec:nonlin_TDSE_gauss}, we
find that the variational TGWD is equivalent to the nonlinear TDSE
(\ref{eq:tdse_q-rep}) with an effective potential (\ref{eq:V_eff}) whose
parameters $V_{0}$, $V_{1}$, and $V_{2}$ satisfy conditions (\ref{eq:V_VGA})
and are all real, as required in Proposition 2. All together, the
Dirac--Frenkel principle applied to the Gaussian (\ref{eq:GWP}) is equivalent
to the system (\ref{eq:q_dot})--(\ref{eq:gamma_dot}) with coefficients $V_{j}$
given by Eq.~(\ref{eq:V_eff}), i.e., to the system (\ref{eq:q_dot_VG}%
)--(\ref{eq:gamma_dot_VG}).

\section{\label{sec:conservation_E_VGA}Conservation of $E_{\text{eff}}$ by the
variational TGWD}

Although $E_{\text{eff}}=E=\operatorname{const}$ for any solution of the
Dirac-Frenkel variational principle (see
Appendix~\ref{sec:cons_E_norm_by_DFVP}), it is instructive to demonstrate the
conservation of the effective energy explicitly for the variational TGWD,
regarded as a solution of the nonlinear TDSE~(\ref{eq:nonlinear_tdse}).
Applying the general expression (\ref{eq:d_exp_Heff_dt}) for $\dot
{E}_{\text{eff}}$ to the variational coefficients~(\ref{eq:V_VGA}) gives%
\begin{align}
\dot{E}_{\text{eff}}  &  =\frac{d}{dt}[\langle\hat{V}\rangle-\operatorname{Tr}%
(V_{2}\cdot\Sigma_{t})/2]-V_{1}^{T}\cdot\dot{q}_{t}+\operatorname{Tr}(\dot
{V}_{2}\cdot\Sigma_{t})/2\nonumber\\
&  =\frac{d\langle\hat{V}\rangle}{dt}-\frac{1}{2}\operatorname{Tr}(V_{2}%
\cdot\dot{\Sigma}_{t})-\langle\hat{V}^{\prime}\rangle^{T}\cdot m^{-1}\cdot
p_{t},
\end{align}
where two terms $\operatorname{Tr}(\dot{V}_{2}\cdot\Sigma_{t})/2$ of opposite
signs canceled each other. The time derivative of $E_{\text{eff}}$ becomes%
\begin{align}
\dot{E}_{\text{eff}}  &  =\operatorname{Tr}(m^{-1}\cdot\operatorname{Re}%
\langle\hat{V}^{\prime}\otimes\hat{p}^{T}\rangle)-\operatorname{Tr}[V_{2}%
\cdot\operatorname{Cov}_{R}(\hat{q},\hat{p})\cdot m^{-1}]\nonumber\\
&  ~~~-\operatorname{Tr}(m^{-1}\cdot\langle\hat{V}^{\prime}\rangle\otimes
p_{t}^{T})
\end{align}
because%
\[
\frac{d\langle\hat{V}\rangle}{dt}=\operatorname{Re}\langle\hat{p}^{T}\cdot
m^{-1}\cdot\hat{V}^{\prime}\rangle=\operatorname{Tr}(m^{-1}\cdot
\operatorname{Re}\langle\hat{V}^{\prime}\otimes\hat{p}^{T}\rangle),
\]
which can be derived in the same way as Eq.~(\ref{eq:comm_H_Heff_1}), and%
\begin{align}
\operatorname{Tr}(V_{2}\cdot\dot{\Sigma}_{t})  &  =\frac{\hbar}{2}%
\operatorname{Tr}[V_{2}\cdot(m^{-1}\cdot\mathcal{A}\cdot\mathcal{B}%
^{-1}+\mathcal{B}^{-1}\cdot\mathcal{A}\cdot m^{-1})]\nonumber\\
&  =2\operatorname{Tr}[V_{2}\cdot\operatorname{Cov}_{R}(\hat{q},\hat{p})\cdot
m^{-1}]. \label{eq:Tr_V2_Sigma_dot}%
\end{align}
To find expression$~$(\ref{eq:Tr_V2_Sigma_dot}) for $\operatorname{Tr}%
(V_{2}\cdot\dot{\Sigma}_{t})$, we used the notation $\mathcal{A}%
:=\operatorname{Re}A_{t}$ and $\mathcal{B}:=\operatorname{Im}B_{t}$ and
relation
\begin{align}
\dot{\Sigma}_{t}  &  =-(\hbar/2)\mathcal{B}^{-1}\cdot\mathcal{\dot{B}}%
\cdot\mathcal{B}^{-1}\nonumber\\
&  =(\hbar/2)(m^{-1}\cdot\mathcal{A}\cdot\mathcal{B}^{-1}+\mathcal{B}%
^{-1}\cdot\mathcal{A}\cdot m^{-1})
\end{align}
which follows from Eqs.~(\ref{eq:Cov_q}) and (\ref{eq:A_dot}).
Equation~(\ref{eq:Tr_V2_Sigma_dot}) then follows from symmetry of matrices
$\mathcal{A}$, $\mathcal{B}^{-1}$, $m^{-1}$, and $V_{2},$ and from
Eq.~(\ref{eq:CovR_qp}) for $\operatorname{Cov}_{R}(\hat{q},\hat{p})$ because%
\begin{align*}
&  \operatorname{Tr}\left(  V_{2}\cdot m^{-1}\cdot\mathcal{A}\cdot
\mathcal{B}^{-1}\right)  =\operatorname{Tr}[\left(  V_{2}\cdot m^{-1}%
\cdot\mathcal{A}\cdot\mathcal{B}^{-1}\right)  ^{T}]\\
&  =\operatorname{Tr}\left(  \mathcal{B}^{-1}\cdot\mathcal{A}\cdot m^{-1}\cdot
V_{2}\right)  =\operatorname{Tr}\left(  V_{2}\cdot\mathcal{B}^{-1}%
\cdot\mathcal{A}\cdot m^{-1}\right)  .
\end{align*}
Using identity (\ref{eq:exp_V_prime_p-pt}) and definition~(\ref{eq:CovR_AB})
of $\operatorname{Cov}_{R}(\hat{q},\hat{p})$ shows that the effective energy
is conserved:%
\begin{align}
\dot{E}_{\text{eff}}  &  =\operatorname{Tr}\{m^{-1}\cdot\lbrack
\operatorname{Re}\langle\hat{V}^{\prime}\otimes(\hat{p}-p_{t})^{T}%
\rangle-\langle\hat{V}^{\prime\prime}\rangle\cdot\operatorname{Cov}_{R}%
(\hat{q},\hat{p})]\}\nonumber\\
&  =0.
\end{align}

\section{\label{app:zero_ReA}Proof that $\mathcal{A}=\operatorname{Re}A=0$ in
FGWD}

Here we prove that setting $\operatorname{Im}V_{2}=0$ in Eq.~(\ref{eq:Im_V2})
for the FGWD implies that the width matrix $A$ of the frozen Gaussian is
purely imaginary. In fact, we will show more generally that if $m^{-1}$ and
$\mathcal{B}$ are positive-definite real symmetric $D\times D$ matrices and
$\mathcal{A}$ a real (but not necessarily symmetric) $D\times D$ matrix, then
the equation
\begin{equation}
\mathcal{A}\cdot m^{-1}\cdot\mathcal{B}+\mathcal{B}\cdot m^{-1}\cdot
\mathcal{A}=0 \label{eq:main}%
\end{equation}
implies that $\mathcal{A}=0$. First, note that matrix $m^{-1}$ has a unique
positive-definite real symmetric square root $m^{-1/2}$%
.\cite{book_Halmos:1942} Multiplication of Eq.~(\ref{eq:main}) by $m^{-1/2}$
both from the left and from the right yields%
\begin{equation}
\mathcal{A}^{\prime}\cdot\mathcal{B}^{\prime}+\mathcal{B}^{\prime}%
\cdot\mathcal{A}^{\prime}=0, \label{eq:main_mass_scaled}%
\end{equation}
where $\mathcal{A}^{\prime}=m^{-1/2}\cdot\mathcal{A}\cdot m^{-1/2}$ is a real
matrix and $\mathcal{B}^{\prime}=m^{-1/2}\cdot\mathcal{B}\cdot m^{-1/2}$ is a
positive-definite real symmetric matrix,\cite{book_Halmos:1942} whose
eigenvectors $v_{j}$ form the basis of $%
\mathbb{R}
^{D}$. In this basis, a matrix element of Eq.~(\ref{eq:main_mass_scaled}) is
\begin{equation}
0=v_{j}^{T}\cdot(\mathcal{A}^{\prime}\cdot\mathcal{B}^{\prime}+\mathcal{B}%
^{\prime}\cdot\mathcal{A}^{\prime})\cdot v_{k}=(\lambda_{j}+\lambda_{k}%
)v_{j}^{T}\cdot\mathcal{A}^{\prime}\cdot v_{k}.\nonumber
\end{equation}
Since all eigenvalues $\lambda_{j}$ of $\mathcal{B}^{\prime}$ are strictly
positive, $v_{j}\cdot\mathcal{A}^{\prime}\cdot v_{k}=0$ for all $j$ and $k$.
Therefore, $\mathcal{A}^{\prime}=0$. Because $m^{-1/2}$ is positive-definite,
it has an inverse $m^{1/2}$ and%
\begin{equation}
\mathcal{A}=m^{1/2}\cdot\mathcal{A}^{\prime}\cdot m^{1/2}=0.
\end{equation}

\section{\label{app:kinetic_propagation_derivation}Derivation of the kinetic
propagation}

Here we derive analytical solutions (\ref{eq:q_t_T_prop}%
)--(\ref{eq:gamma_t_T_prop}) for the kinetic propagation. Equations
(\ref{eq:q_t_T_prop})\ and (\ref{eq:p_t_T_prop}) for $q_{t}$ and $p_{t}$
follow from Eqs.~(\ref{eq:q_dot_T_prop}) and (\ref{eq:p_dot_T_prop}) because
$p_{t}=\operatorname{const}$. To solve the differential
Eq.~(\ref{eq:A_dot_T_prop}) for $A_{t}$, we use the relation for the
derivative of a matrix inverse:\cite{Petersen_Pedersen:2012}%
\begin{equation}
d(A_{t}^{-1})/dt=-A_{t}^{-1}\cdot\dot{A}_{t}\cdot A_{t}^{-1}=m^{-1},
\label{eq:dAt_inv}%
\end{equation}
where Eq.~(\ref{eq:A_dot_T_prop}) was used in the second step. Differential
equation (\ref{eq:dAt_inv})\ has a trivial solution%
\begin{align}
A_{t}^{-1}  &  =A_{0}^{-1}+tm^{-1}=A_{0}^{-1}\cdot(\operatorname{Id}%
_{D}+tA_{0}\cdot m^{-1})\nonumber\\
&  =(\operatorname{Id}_{D}+tm^{-1}\cdot A_{0})\cdot A_{0}^{-1},
\label{eq;At_inverse_from_A0_inverse}%
\end{align}
and taking the inverse of the three alternative expressions for $A_{t}^{-1}$
gives the three formulas for $A_{t}$ in Eq.~(\ref{eq:A_t_T_prop}). To find
$\gamma_{t}$, we substitute the expression for $m^{-1}\cdot A_{t}$ from
Eq.~(\ref{eq:A_dot_T_prop}) into Eq.~(\ref{eq:gamma_dot_T_prop}) and use the
formula for the derivative of a logarithm of a determinant:%
\begin{align}
\dot{\gamma}_{t}  &  =T(p_{t})-(i\hbar/2)\operatorname*{Tr}(A_{t}^{-1}%
\cdot\dot{A}_{t})\nonumber\\
&  =T(p_{t})-(i\hbar/2)\frac{d}{dt}\ln\det A_{t}.
\label{eq:gamma_dot_T_prop_1}%
\end{align}
Since $p_{t}=p_{0}$, this differential equation has the solution%
\begin{align}
\gamma_{t}  &  =\gamma_{0}+tT(p_{0})-(i\hbar/2)\ln[(\det A_{t})/(\det
A_{0})]\nonumber\\
&  =\gamma_{0}+tT(p_{0})-(i\hbar/2)\ln\det(A_{0}^{-1}\cdot A_{t}).
\label{eq:gamma_t_T_prop_1}%
\end{align}
Inserting Eq.~(\ref{eq:A_t_T_prop}) for $A_{t}$ gives
Eq.~(\ref{eq:gamma_t_T_prop}) for $\gamma_{t}$.

Among Eqs.~(\ref{eq:q_t_T_prop})--(\ref{eq:gamma_t_T_prop}), the only one that
is nontrivial to invert is Eq.~(\ref{eq:gamma_t_T_prop}) for $\gamma_{t}$,
whose inversion (\ref{eq:gamma_0_inv_T_prop}) follows from
Eqs.~(\ref{eq:q_t_T_prop}) and (\ref{eq;At_inverse_from_A0_inverse}) because%
\begin{align}
&  \operatorname{Id}_{D}+tm^{-1}\cdot A_{0}=A_{t}^{-1}\cdot A_{0}=(A_{0}%
^{-1}\cdot A_{t})^{-1}\nonumber\\
&  =[(A_{t}^{-1}-tm^{-1})\cdot A_{t}]^{-1}=\left(  \operatorname{Id}%
_{D}-tm^{-1}\cdot A_{t}\right)  ^{-1}.
\end{align}

\bibliographystyle{aipnum4-2}
\bibliography{nonlinear_GWP_v60}

\end{document}